# Single-atom catalysis in space: Computational exploration of Fischer–Tropsch reactions in astrophysical environments[★]

G. Pareras[1], V. Cabedo[2], M. McCoustra[2], and A. Rimola[1]

[1] Departament de Química, Universitat Autònoma de Barcelona, 08193 Bellaterra, Catalonia, Spain
  e-mail: albert.rimola@uab.cat; gerard.pareras@uab.cat
[2] Institute of Chemical Sciences, School of Engineering and Physical Sciences, Heriot-Watt University, Edinburgh EH14 4AS, Scotland, UK

**ABSTRACT**

*Context.* Gas-phase chemistry at extreme conditions (low densities and temperatures) is difficult, so the presence of interstellar grains is especially important for the synthesis of molecules that cannot form in the gas phase. Interstellar grains are advocated to enhance the encounter rate of the reactive species on their surfaces and to dissipate the energy excess of largely exothermic reactions, but less is known of their role as chemical catalysts that provide low activation energy pathways with enhanced reaction rates. Different materials with catalytic properties are present in interstellar environments, like refractory grains containing space-abundant *d*-block transition metals.
*Aims.* In this work we report for first time mechanistic insights on the Fischer–Tropsch methanol ($CH_3OH$) synthesis under astrophysical conditions using single-atom Fe-containing silica surfaces as interstellar heterogeneous catalysts.
*Methods.* Quantum chemical calculations considering extended periodic surfaces were carried out in order to search for the stationary points and transitions states to finally construct the reaction potential energy surfaces. Binding energy and kinetic calculations based on the Rice–Ramsperger–Kassel–Marcus (RRKM) scheme were also performed to evaluate the catalytical capacity of the grain and to allocate those reaction processes within the astrochemical framework.
*Results.* Our mechanistic studies demonstrate that astrocatalysis is feasible in astrophysical environments. Thermodynamically the proposed process is largely exergonic, but kinetically it shows energy barriers that would need from an energy input in order to go through. Kinetic calculations also demonstrate the strong temperature dependency of the reaction process as tunnelling is not relevant in the involved energetic barriers. The present results can explain the presence of $CH_3OH$ in diverse regions where current models fail to reproduce its observational quantity.
*Conclusions.* The evidence of astrocatalysis opens a completely new spectrum of synthetic routes triggering chemical evolution in space. From the mechanistic point of view the formation of methanol catalysed by a single atom of $Fe^0$ is feasible; however, its dependency on the temperature makes the energetics a key issue in this scenario.

## 1. Introduction

The detection of organic molecules in the interstellar medium (ISM), specifically in regions that will eventually form solar-like planetary systems, is one of the most remarkable discoveries of modern astronomy. The presence of these interstellar complex organic molecules (iCOMs) sets a direct link between organic chemistry and solar-type systems (Ceccarelli et al. 2017; Herbst & Van Dishoeck 2009; Saladino et al. 2012; Ceccarelli 2000). Moreover, since some of these species have prebiotic potential (e.g. formamide, $NH_2CHO$; Snyder et al. 1969), iCOMs are considered the small bricks from which the macromolecules of terrestrial life are assembled. Life on Earth likely arose from a series of conditions that remain to be unambiguously determined. The early Earth was exposed not only to different sources of energy, but also to a large flux of extra-terrestrial material (Bottke et al. 2012). Atmospheric and surface conditions of the early Earth promoted an increase in the molecular complexity allowing the formation of more complex, interacting and self-replicating systems (Rotelli et al. 2016). Thus, the iCOM small bricks are understood as the chemical seeds of life.

The synthesis of iCOMs has been extensively studied in order to reveal the evolution of interstellar chemistry, that is, to understand the emergence of complex chemistry in progenitor solar-like systems, with the goal of knowing how organic chemistry builds up and evolves and, ultimately, how universal these chemical seeds of life are (Herbst 2017). The prevailing paradigms for the iCOMs synthesis postulates that they form on the surfaces of interstellar grains (Garrod et al. 2008). The simplest organic compounds, formaldehyde ($CH_2O$) and methanol ($CH_3OH$), have been repeatedly observed from starless cores to star-forming regions (Van Gelder et al. 2020; McGuire 2018; Bianchi et al. 2019; López-Sepulcre et al. 2015; Vasyunina et al. 2014); the widely accepted scenario is that they form by the hydrogenation of CO by the successive addition of H atoms occurring on icy grains. Experimental (Hiraoka et al. 2002; Watanabe & Kouchi 2002; Watanabe et al. 2003; Hidaka et al. 2004; Fuchs et al. 2009; Pirim et al. 2010; Minissale et al. 2016; Ohno et al. 2022) and theoretical (Rimola et al. 2014;

---

[★] The data supporting this article are freely available at Zenodo at https://doi.org/10.5281/zenodo.8380174

Korchagina et al. 2017; Song & Kästner 2017; Kuwahata & Ohno 2018; Pantaleone et al. 2020; Enrique-Romero et al. 2021; Tieppo et al. 2023) evidence support this reaction formation mechanism.

Dust grain particles are ubiquitously present at different astrophysical evolutionary stages. They are crucial players in the chemical evolution in space as they provide surface sites, which help the formation of some simple molecules and iCOMs (Zamirri et al. 2019; Öberg 2016; Linnartz et al. 2015; Watanabe & Kouchi 2008; Vidali 2013; Hama & Watanabe 2013; Van Dishoeck et al. 2013; Wakelam et al. 2017). Grains consist of a core of refractory materials, mainly silicates in O-rich stars (Henning 2010). At the low temperatures of the ISM in prestellar cores (10 K), most volatiles are condensed on the surface of the grains forming ice mantles. However, recent findings indicate that, due to the porous structure of the grain cores, ices do not necessarily fully cover them, leaving the refractory materials partially exposed to the environmental gas (Potapov et al. 2020).

The underlying idea behind the grain chemistry is that some chemical reactions are favoured when they occur on the grain surfaces. However, state-of-the-art investigations mainly focus on the role of grains in concentrating the reactants and acting as third bodies by absorbing the energy excess of the reactions. Moreover, they are normally focused on ices, and not on the grain surface themselves. Consequently, investigations into actual grain chemical catalysis, namely grain surfaces providing alternative routes with lower energy barriers, are practically absent.

The catalytic efficiency of the grains strongly depends on its atomistic features, in particular the presence and nature of surface defects or possible active sites. Catalytically active species, such as iron (Fe), are relatively abundant in space (Lodders 2003) and can be incorporated in solid phases as they condense from the gas phase, giving rise to single-atom catalytic systems (Adams & Mccord 1971; Nuth et al. 2006). Additionally, space weathering and sputtering of silicates can give rise to material rich in silica ($SiO_2$; Sargent et al. 2009) with Fe inclusions (Sasaki et al. 2001). Therefore, the likely presence of heterogeneous catalysts in the ISM opens a new perspective on interstellar synthesis. Moreover, the omnipresence of these catalytic systems makes it possible to extrapolate this reactivity beyond the (cold) first steps of the star formation (e.g. protoplanetary disks or planetary atmospheres), where temperatures and densities are higher.

Since $H_2$ and CO are the two most abundant gas-phase compounds in most dense astrophysical environments, and considering that Fe is the most space-abundant transition metal element, Fischer–Tropsch-type (FTT) catalytic processes can likely operate in the ISM. We note that transition metals here indicate those elements that belong to the d-block of the periodic table. Fischer-Tropsch reactions are the current industrial synthesis of hydrocarbons and alcohols from mixing $H_2$ and CO, which is favoured and accelerated in the presence of a metal catalyst, typically Fe, but also Ni, Co, Ru, or Cr. The formation of hydrocarbons (alkanes and alkenes) and alcohols through FTT processes occur via the following reactions:

$$n\text{CO} + (2n+1)\text{H}_2 \rightleftharpoons \text{C}_n\text{H}_{2n+2} + n\text{H}_2\text{O}, \qquad (1)$$

$$n\text{CO} + (2n)\text{H}_2 \rightleftharpoons \text{C}_n\text{H}_{2n} + n\text{H}_2\text{O}, \qquad (2)$$

$$n\text{CO} + (2n)\text{H}_2 \rightleftharpoons \text{C}_n\text{H}_{2n+1}\text{OH} + (n-1)\text{H}_2\text{O}. \qquad (3)$$

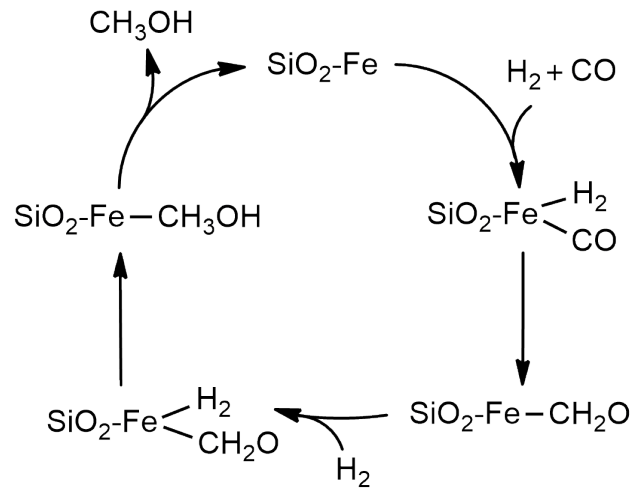

**Fig. 1.** Proposed catalytic cycle for methanol ($CH_3OH$) formation from the reactivity of $H_2$ and CO adopting the Fischer–Tropsch-type mechanism, in which the catalyst is a single atom of Fe attached on a silica ($SiO_2$) surface.

Accordingly, the occurrence of FTT in space would mean the establishment of a significant, additional synthetic route towards iCOMs based on chemical catalysis in space. FTT reactions have been tested experimentally on icy and metal-containing grain analogues and meteoritic material under diverse space conditions. These experiments lead to the formation of small (C1–C4) hydrocarbons, aromatic hydrocarbons, and simple alcohols, such as methanol ($CH_3OH$; Rotelli et al. 2016; Llorca & Casanova 2000; Cabedo et al. 2021; Qasim et al. 2020). However, comprehensive studies rationalising these findings (e.g. product distributions, mechanistic steps, rate measurements, and catalytic metal dependence) and revealing the actual catalytic activity of the grains are missing. Interestingly, the underlying idea of true astrocatalysis has been put forward in Fioroni & DeYonker (2022), by studying theoretically the hydrocyanation of alkenes catalysed by Ni centres (the second most space-abundant transition metal after Fe) on siliceous surfaces.

Here we present the first computational study, based on quantum chemical simulations, of FTT reactions expected to occur on ISM dust grain surfaces. The target is the simplest FTT reaction, namely the production of $CH_3OH$ from two $H_2$ and one CO molecule catalysed by Fe centres, adopting the catalytic cycle shown in Fig. 1. The mechanistic steps, the related energetics, and the catalytic efficiency of this process have been simulated, considering Fe single atoms ($Fe^0$ and $Fe^{2+}$) supported on $SiO_2$ surfaces ($Fe^0@SiO_2$ and $Fe^{2+}@SiO_2$) as the catalytic systems. Based on the kinetic calculations, a final discussion on the astrophysical implications of our results is given by proposing in which regions the investigated FTT processes are feasible, suggesting that their occurrence can alleviate the disagreement between observations and model predictions on the $CH_3OH$ abundances in some astrophysical environments.

## 2. Methodology

### 2.1. Quantum chemical simulations

Quantum chemical calculations were carried out within the periodic boundary conditions (PBC) framework. Geometry optimisations of the periodic systems were carried out with the CP2K package (Hutter et al. 2014). The semi-local generalised gradient approximation (GGA) Perdew–Burke–Ernzerhof (PBEsol) functional was adopted (Perdew et al. 2008) together with the Grimme D3(BJ) correction to include dispersion forces (Grimme et al. 2010), combined with a double-$\zeta$ basis set (DZVP-MOLOPT-SR-GTH Gaussian basis set) for all the atom types, and a cut-off set at 500 Ry for the plane wave auxiliary basis set. Core electrons were described with the Goedecker–Teter–Hutter pseudopotentials (Goedecker & Teter 1996) and valence electrons with a mixed Gaussian and plane-wave (GPW) approach (Lippert et al. 1997).

To refine the energies of the stationary points, single point calculations on the optimised geometries were performed employing the Becke-Lee, Yang, Parr hybrid GGA (B3LYP) functional (Becke 1993; Lee et al. 1988; Stephens et al. 1994), including the Grimme D3(BJ) correction term (Grimme et al. 2010), and using the triple-$\zeta$ basis set (TZVP) with the ab initio CRYSTAL17 code (Dovesi et al. 2018). The threshold parameters for the evaluation of the Coulomb and exchange bielectronic integrals (TOLINTEG keyword in the CRYSTAL17 code; Dovesi et al. 2018) were set to 7, 7, 7, 7, and 14. B3LYP was selected to perform these energy refinement calculations because a preliminary benchmark study demonstrated this functional to be the best one to describe the electronic structure of the different electronic states of our systems (see Appendix A).

Transition states have been searched for using the Climbing Image Nudged Elastic Band (CI-NEB) technique implemented in CP2K (Hutter et al. 2014). Activation energy barriers have been calculated as

$$\Delta E^{\ddagger} = E_{TS} - E_{GS}, \quad (4)$$

$$\Delta U^{\ddagger} = \Delta E^{\ddagger} + \Delta^{m} ZPE, \quad (5)$$

$$\Delta G_T^{\ddagger} = \Delta E^{\ddagger} + \Delta G_T. \quad (6)$$

Here $\Delta E^{\ddagger}$ is the potential energy barrier, where $E_{TS}$ and $E_{GS}$ are the absolute potential energies for the transition state and the local minimum structure, respectively; $\Delta U^{\ddagger}$ is the vibrational zero-point energy (ZPE) corrected barrier, where $\Delta ZPE$ refers to the contribution of the ZPE corrections to $\Delta E^{\ddagger}$; and $\Delta G_T^{\ddagger}$ is the Gibbs activation barrier at a given temperature, where $\Delta G_T$ refers to the contribution of the Gibbs corrections to $\Delta E^{\ddagger}$.

Binding energies were computed applying the counterpoise method as implemented in CRYSTAL17 in order to avoid the basis set superposition error (BSSE; Dovesi et al. 2018). The final binding energies were calculated as

$$\Delta E_{ads} = E_{cplx} - (E_{sur} + E_m), \quad (7)$$

$$\Delta U_{ads} = \Delta E_{ads} + \Delta ZPE, \quad (8)$$

$$\Delta U_{ads} = -BE, \quad (9)$$

where $\Delta E_{ads}$ is the potential adsorption energy; $E_{cplx}$, $E_{sur}$, and $E_m$ are the absolute potential energies for the adsorption complex, the isolated surface, and the isolated molecule, respectively; $\Delta U_{ads}$ is the ZPE-corrected adsorption energy (with $\Delta ZPE$ the contribution of the ZPE corrections to $\Delta E_{ads}$); and BE is the binding energy (which is $\Delta U_{ads}$ in opposite sign).

Frequency calculations were performed to validate the nature of the stationary points (i.e. minima with only real frequencies, and transition states with only one imaginary frequency leading to the corresponding product). Vibrational harmonic frequencies were calculated at the PBE-D3BJ/DZVP-optimised structures using the finite differences method as implemented in the CP2K code. A partial Hessian approach was used to reduce the computational cost of the calculations. Thus, the vibrational frequencies were calculated only for a fragment of the entire system, which included the metal centre, the reactive species, and the closest $SiO_2$ moieties of the silica surface.

### 2.2. Kinetic calculations

The rate constants associated with each elementary barrier was calculated using the Rice–Ramsperger–Kassel–Marcus (RRKM) theory (Marcus 1952), a microcanonical version of Eyring's transition state theory (TST), in which all the TST assumptions hold, assuming additionally that the phase space is statistically populated.

For a reaction step with an energy barrier $E_0$, the microcanonical rate constant $k(E)$ at a given total energy $E$ (with $E > E_0$) is given by

$$k(E) = \frac{N^{\neq}(E - E_0)}{hp(E)} = \frac{\int_0^{E_0} T(\xi^{\neq}) d\xi^{\neq} \int_0^{E-E_0} \rho^{\neq}(E - E_0 - \xi^{\neq}) d\xi^{\neq}}{hp(E)}, \quad (10)$$

where $N^{\neq}(E - E_0)$ is the sum of states for the TS with energies from 0 to $E - E_0$, $\rho(E)$ is the density of states of reactants (i.e. the number of states per unit of energy, in quantum mechanics the degeneracy $g(E)$), and $h$ the Planck constant. The sum of states of the TS can be expressed as an integral of its density of states over the translational energy of the reaction coordinate $\xi^{\neq}$. Tunnelling effects are included by introducing the transmission probability $T(\xi^{\neq})$ inside the integral, where the integration range expands from 0 to $E_0$. In a final step, microcanonical $k(E)$ can be converted into canonical $k(T)$ by populating the energy levels according to the Boltzmann distribution, $P(E)$

$$k(T) = \int_{E_0}^{\infty} k(E) x P(E) dE = \int_{E_0}^{\infty} k(E) x \frac{g(E) exp(-E/k_B T)}{Q(T)} dE, \quad (11)$$

where $Q(T)$ is the total partition function.

Analytical expressions for $T(\xi^{\neq})$ can be obtained by using certain potential barrier shapes. Here the unsymmetrical Eckart potential barrier was used (Eckart 1930), as it has been proved to be accurate enough to account for tunnelling in reactions of astrochemical interest (Lamberts et al. 2016).

Finally, all the rate constants are represented in Arrhenius format (i.e. plots of $\ln[k(T)]$ vs $1/T$) as the effect of temperature on the rate constants is given by

$$\ln[k(T)] = -\frac{E_a}{R}\left(\frac{1}{T}\right) + \ln(A), \quad (12)$$

where $k$ is the rate constant, $E_a$ denotes the activation energy, $A$ is the pre-exponential factor, $R$ is the ideal gas constant, and $T$ is the temperature. We note that a change in the slope in an Arrhenius plot indicates the dominance of tunnelling effects.

### 2.3. Surface models for Fe single atoms on $SiO_2$

The heterogeneous catalytic systems were modelled by adsorbing either $Fe^0$ or $Fe^{2+}$ species on silica surfaces (hereafter referred to as $Fe^0@SiO_2$ and $Fe^{2+}@SiO_2$, respectively).

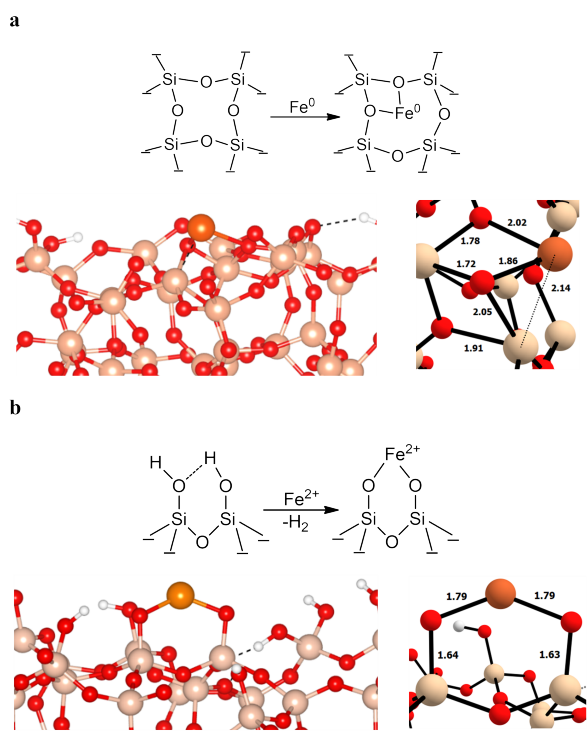

**Fig. 2.** Surface models for Fe single atoms on $SiO_2$. (a) Schematic of $Fe^0$ grafting over the $SiO_2$; shown is a zoomed-out (left) and a zoomed-in (right) image for the optimised geometry of the $Fe^0$@$SiO_2$ system. (b) Schematic of the $Fe^{2+}$ grafting over the $SiO_2$; shown is a zoomed-out (left) and a zoomed-in (right) image for the optimised geometry of the $Fe^{2+}$@$SiO_2$ system. The relevant distances are depicted in angstroms (Å). Colour-coding: white, H atoms; grey, C atoms; red, O atoms; beige, Si atoms; and orange, Fe atoms.

Figure 2 shows the resulting structures. It is worth mentioning that, in both Fe species, only their respective electronic ground states (the most stable states) were considered, which correspond to the high spin states, that is, $Fe^0$ a triplet state (with two unpaired electrons) and $Fe^{2+}$ a quintet state (with four unpaired electrons). The adsorption of the Fe atoms was done on two different amorphous $SiO_2$ models taken from the work of Ugliengo et al. (2008), differing on the silanol (SiOH) surface densities: 1.5 OH nm$^{-2}$ and 4.5 OH nm$^{-2}$. This choice is due to the different nature of the Fe centres in terms of oxidation states (namely $Fe^0$ and $Fe^{2+}$). This means that the surface with 1.5 OH nm$^{-2}$ densities contains $Fe^0$ because it can be adsorbed through the formation of two Fe–O bonds with siloxane (Si–O–Si) groups. In contrast, $Fe^{2+}$ can be grafted to the surface with 4.5 OH nm$^{-2}$ densities by forming two Fe–O bonds belonging to SiOH groups, which are deprotonated, hence compensating for the resulting negative charges. The generated models also served to investigate how Fe single atoms interact and can be found on $SiO_2$ surfaces. Interestingly, in $Fe^0$@$SiO_2$, the two Fe-O bonds are slightly different (2.02 and 1.86 Å; see Fig. 2), and a surface Si atom establishes a slight interaction with the Fe atom, with a Fe-Si distance of 2.14 Å (inducing an increase in the Si–O length, 2.05 Å). Here a charge donation from the filled Fe d orbitals to the empty Si d orbitals can explain this Fe···Si interaction, where the charge transfer allows a hypervalent expansion of the Si atom. In contrast, in $Fe^{2+}$@$SiO_2$, the two Fe–O bonds are equal (and shorter, 1.79 Å) due to classical electrostatic interactions.

## 3. Results

The stationary points and the potential energy surfaces (PESs) for the reactions occurring on $Fe^0$@$SiO_2$ and $Fe^{2+}$@$SiO_2$ are depicted in Figs. 3 and 4, respectively. The figures also present the energy profiles including ZPE corrections and the free energies at 200, 400, and 600 K, and thus considering temperatures of different astrophysical environments at which the astrocatalysts can be found (namely prestellar cores, protostellar envelopes, protoplanetary disks, and planetary atmospheres). For the sake of clarity, we first focus on the ZPE-corrected values and afterwards on free energies to assess the temperature effects. The optimised geometries of all the stationary points can be found in Figs. D.1 and D.2.

The proposed catalytic route (Fig. 1) starts with the adsorption of $H_2$ and CO on the Fe single atoms. We considered that $H_2$ is first adsorbed because it is more abundant than CO in the ISM. However, the opposite situation (namely CO adsorption followed by $H_2$ adsorption) does not significantly modify the stability of the adsorption complexes (see Table C.1). The next step deals with the dissociation of the adsorbed $H_2$ and the transfer of the two H atoms to CO, forming formaldehyde ($CH_2O$). Then a second $H_2$ molecule is adsorbed, and both H atoms transfer to $CH_2O$ forming methanol ($CH_3OH$).

### 3.1. FTT reaction on $Fe^0$@$SiO_2$

On $Fe^0$@$SiO_2$, the adsorption of $H_2$ and CO is exergonic, giving rise to a complete adsorption complex of 60.6 kcal mol$^{-1}$ below the asymptote (Fig. 3b). Interestingly, CO adsorption induces a homolytic H–H bond cleavage, which is assisted by the surface, as each H atom attaches to the $Fe^0$ centre and a surface Si atom (structure C). This spontaneous dissociation is key to the process because starting with an already dissociated $H_2$ molecule facilitates the CO hydrogenations. As mentioned before, on the bare $Fe^0$@$SiO_2$ surface, the $Fe^0$ and Si atoms establish a weak Fe···Si interaction. The first H transfer ($TS_{C-D}$) presents an intrinsic ZPE-corrected energy barrier of 21.6 kcal mol$^{-1}$. Gibbs free energies show a stabilisation of $TS_{C-D}$, which reduces the barrier to 15.8 kcal mol$^{-1}$ at 600 K. This step leads to the endergonic formation of the D intermediate. The second H transfer ($TS_{D-E}$) presents a ZPE-corrected energy barrier of 15.16 kcal mol$^{-1}$, which also becomes reduced to 10.30 kcal mol$^{-1}$ when considering Gibbs free energies at 600 K. This second hydrogenation leads to the exergonic formation of the intermediate $E$, corresponding to the $CH_2O$ formation. The reaction proceeds with the incorporation of a second $H_2$ molecule (structure F), with an adsorption energy of around –20 kcal mol$^{-1}$. Here the spontaneous $H_2$ dissociation is not observed, and $H_2$ coordinates the $Fe^0$ centre through its H–H bond. The third and fourth H transfers ($TS_{F-G}$ and $TS_{G-H}$) to form $CH_3O$ and the final $CH_3OH$ (structures G and H), respectively, present ZPE-corrected energy barriers of 12.7 and 17.0 kcal mol$^{-1}$. In these cases an increase in temperature does not significantly affect the barrier heights (being almost the same at 600 K), and both steps are exergonic.

### 3.2. FTT reaction on $Fe^{2+}$@$SiO_2$

On $Fe^{2+}$@$SiO_2$, the adsorption of $H_2$ and CO is favourable (less, however, than on $Fe^0$@$SiO_2$ probably because there is less electron donation from $Fe^{2+}$ to the adsorbates than from $Fe^0$). The resulting structure is the adsorption complex C situated at –35.83 kcal mol$^{-1}$ with respect to the asymptote (Fig. 4b). The first H transfer occurs through TS$_{C-D}$ with a ZPE-corrected barrier of 54.87 kcal mol$^{-1}$, which leads to the endergonic formation of D. This energy barrier is considerably higher than that on Fe$^{2+}$@SiO$_2$ because, in this case, H$_2$ does not spontaneously dissociate upon adsorption, so this step mainly deals with the H-H bond cleavage. At variance with the Fe$^0$@SiO$_2$ analogue step, a temperature increase does not modify the energetics. The second hydrogen transfer is barrierless by introducing ZPE corrections, forming CH$_2$O in E in an exergonic way. The adsorption of the second H$_2$ (structure F) does not confer significant stability to the system (4 kcal mol$^{-1}$ more stable than structure E) because no charge transfer from Fe$^{2+}$ to H$_2$ occurs (this already happens with CH$_2$O), and so the interaction is essentially through dispersion. This non-activation of the H$_2$ implies that the third hydrogenation (TS$_{F-G}$) proceeds with a ZPE-corrected barrier of 22.1 kcal mol$^{-1}$ (22.4 kcal mol$^{-1}$ when considering free energies at 600 K), which is about 10 kcal mol$^{-1}$ higher than on Fe$^0$@SiO$_2$; moreover, the formation of CH$_3$O (structure G) is endergonic. However, the last H transfer (TS$_{G-H}$) presents very low ZPE-corrected (and free energy) barriers (3.6 kcal mol$^{-1}$), leading to the exergonic formation of the CH$_3$OH (structure H).

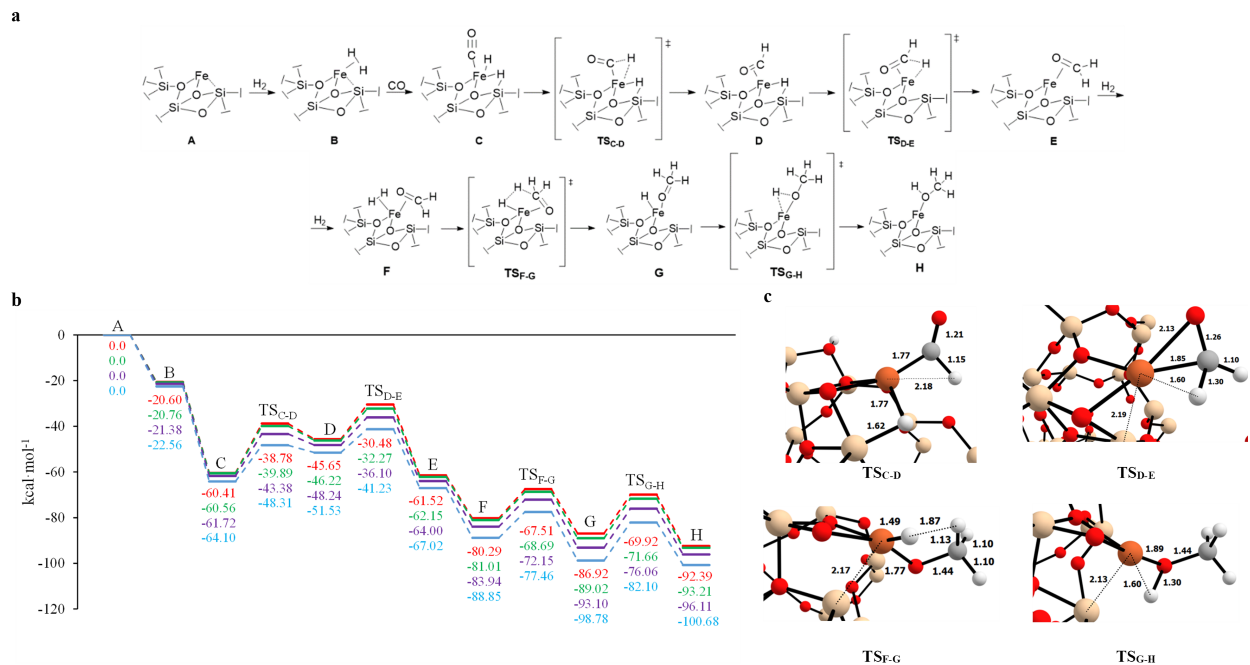

**Fig. 3.** FTT-based reactivity for the CH$_3$OH formation on Fe$^0$@SiO$_2$. (a) Reaction mechanism identified for CH$_3$OH formation on Fe$^0$@SiO$_2$. (b) Energy profiles (in kcal mol$^{-1}$) of CH$_3$OH formation, using as 0th reference energy the A + H$_2$ + CO asymptote. Colour-coding: red, relative ZPE-corrected energies; green, purple, and blue, relative Gibbs energies at 200, 400, and 600 K, respectively. (c) Optimised geometries of the four transition state structures (bond distances in Å). Colour-coding: white, H atoms; grey, C atoms; red, O atoms; beige, Si atoms; and orange, Fe atoms.

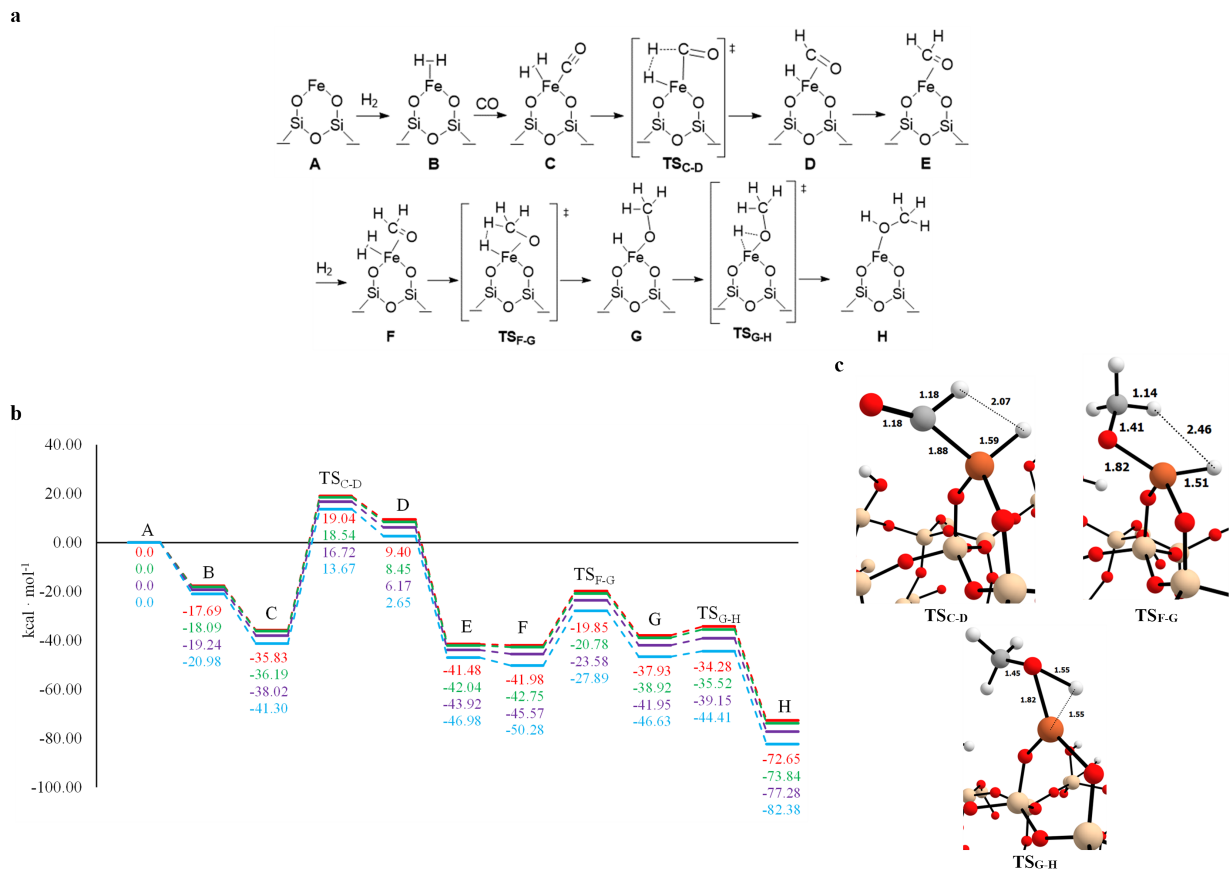

**Fig. 4.** FTT-based reactivity for the CH$_3$OH formation on Fe$^{2+}$@SiO$_2$. (a) Reaction mechanism identified for CH$_3$OH formation on Fe$^{2+}$@SiO$_2$. (b) Energy profiles (in kcal mol$^{-1}$) of the CH$_3$OH formation, using as 0th reference energy the A + H$_2$ + CO asymptote. Colour-coding: red, relative ZPE-corrected energies; green, purple, and blue, relative Gibbs energies at 200, 400, and 600 K, respectively. (c) Optimised geometries of the four transition state structures (bond distances in Å). Colour-coding: white, H atoms; grey, C atoms; red, O atoms; beige, Si atoms; and orange, Fe atoms.

## 4. Discussion

### 4.1. Energetic implications: Binding energies and kinetic analysis

Although the mechanistic steps are similar, and formation of CH$_2$O and CH$_3$OH on both catalysts are thermodynamically favourable, important differences are found in relation to the energy barriers. In both systems the step presenting the highest ZPE-corrected energy barrier is the first hydrogenation, but on Fe$^0$@SiO$_2$ the barrier is of ≈22 kcal mol$^{-1}$, while on Fe$^{2+}$@SiO$_2$ ≈55 kcal mol$^{-1}$: in the former case, the first H$_2$ is homolytically dissociated, while this is not happening in the latter case, indicating the major catalytic activity of the Fe$^0$@SiO$_2$ surface. Moreover, on Fe$^0$@SiO$_2$ temperature induces a reduction of the free energy barrier to ≈15 kcal mol$^{-1}$, while on Fe$^{2+}$@SiO$_2$ temperature hardly affects the energetics.

This energy difference between the two catalysts is also important regarding the overall energy profiles. That is, on Fe$^0$@SiO$_2$ all the identified transitions states are lower in energy than the asymptote A and of the H$_2$/Fe$^0$@SiO$_2$ (intermediate B) complex (Fig. 3b). In contrast, on Fe$^{2+}$@SiO$_2$, TS$_{C-D}$ and intermediate D are higher in energy than the asymptote A (Fig. 4b). This means that, on Fe$^0$@SiO$_2$, the energy released by the CO adsorption, if it is not quickly dissipated throughout the surface, can be used to overcome all the elementary steps, while this cannot take place on Fe$^{2+}$@SiO$_2$. Accordingly, on Fe$^0$@SiO$_2$ the reaction can proceed favourably in an apparent barrierless way, while on Fe$^{2+}$@SiO$_2$ it is hindered by the presence of an actual energy barrier.

It is worth mentioning, however, that in this case where the adsorption energies are not dissipated throughout the surface, it is possible that the same retained energy can induce the desorption of reaction intermediates (like CH$_2$O), in this way interrupting the progress of the reaction. To further understand this aspect, we calculated the BEs of the reactants, intermediates, and products on the catalyst surfaces (see Table 1) and compared them with the energetics of the reactions. In general, the BEs of all these species on the catalyst models show a strong interaction between the two partners. Moreover, if we focus on the intermediate and product (CH$_2$O and CH$_3$OH, respectively), BEs provide additional information of interest. In the

**Table 1.** Collected binding energies (BEs).

| System | BE H$_2$ | BE CO | BE CH$_2$O | CH$_3$OH |
|---|---|---|---|---|
| Fe$^0$@SiO$_2$ | 34.58 | 41.39 | 74.02 | 58.39 |
| Fe$^{2+}$@SiO$_2$ | 23.78 | 22.86 | 48.99 | 41.67 |

**Notes.** Energies are in kcal mol$^{-1}$ and BEs calculated here are between the adsorbates (CO and H$_2$), intermediate (CH$_2$O) and product (CH$_3$OH), and each SiO$_2$ surface under study.

case of Fe$^0$@SiO$_2$, the BE of CH$_2$O is 74.02 kcal mol$^{-1}$, while the intermediate step (structure E of Fig. 3b) is located at −61.52 kcal mol$^{-1}$ with respect to the asymptote. As the energy required to desorb the CH$_2$O molecule is located above the asymptote, this means that CH$_2$O is doomed to remain attached on the surface, allowing the reaction to continue. On the other hand, the BE for the CH$_3$OH molecule is 58.39 kcal mol$^{-1}$, while the final step (structure H of Fig. 3b) is at −92.39 kcal mol$^{-1}$ with respect to the asymptote. Here the BE is allocated below the asymptote, and therefore, once the CH$_3$OH is formed, it could be released to the gas-phase. On Fe$^{2+}$@SiO$_2$, a similar behaviour is observed, although the BEs are, in general, smaller than those on Fe$^0$@SiO$_2$.

On the other hand, in the case where surfaces can act as effective third bodies by absorbing and dissipating the energies released by exothermic processes, all the intermediate species would be thermalised systems that need to overcome the energy barriers that connect them. To have deeper insights into the feasibility of the reactions within this scenario, we carried out kinetics calculations, including tunnelling, of all the elementary steps to derive their rate constants at the considered temperatures.

Figure 5 represents the Arrhenius plots associated with the rate determining steps (RDS) of the FTT processes on Fe$^0$@SiO$_2$ and Fe$^{2+}$@SiO$_2$. Arrhenius plots associated with the rest of elementary energy barriers are included in Figs. F.1 and F.2. For the sake of the astrochemical context, the units are in yr$^{-1}$.

On Fe$^0$@SiO$_2$ (see Figs. 5a and F.1), the derived Arrhenius activation energies are in good agreement with the ZPE-corrected energies (22.16 kcal mol$^{-1}$, 15.47 kcal mol$^{-1}$, 13.01 kcal mol$^{-1}$, and 17.39 kcal mol$^{-1}$, for TS$_{C-D}$, TS$_{D-E}$, TS$_{F-G}$, and TS$_{G-H}$, respectively). Therefore, kinetic data confirms that the rate determining step (RDS) is the first energetic barrier. As it is endergonic, tunnelling contributions can be ruled out and the kinetics totally depend on the temperature. The two following steps (Figs. F.1b and F.1c) are exergonic. Tunnelling presents a contribution, but a very minor one and only at very low temperatures, and thus, in practice, the rates can be considered very low. In contrast, in the last step, tunnelling dominates at low temperatures (the change in slope begins at 200 K) and below 100 K the reaction is in the tunnelling regime, but still presents very low rates. Therefore, these data indicate that the reaction is a temperature-dependent process and that it cannot proceed in the coldest regions of the ISM. However, focusing on the RDS, the rate constant is 1 yr$^{-1}$ when the temperature is around 220 K, but due to the strong temperature dependence, at 250 K it significantly increases to 626 yr$^{-1}$; we note that the reaction can show a catalytic enhancement in the later stages of the planetary system formation (with higher temperatures than the interstellar clouds).

On Fe$^{2+}$@SiO$_2$ (see Figs. 5b and F.2), the Arrhenius activation energies derived from the Arrhenius plots are in good

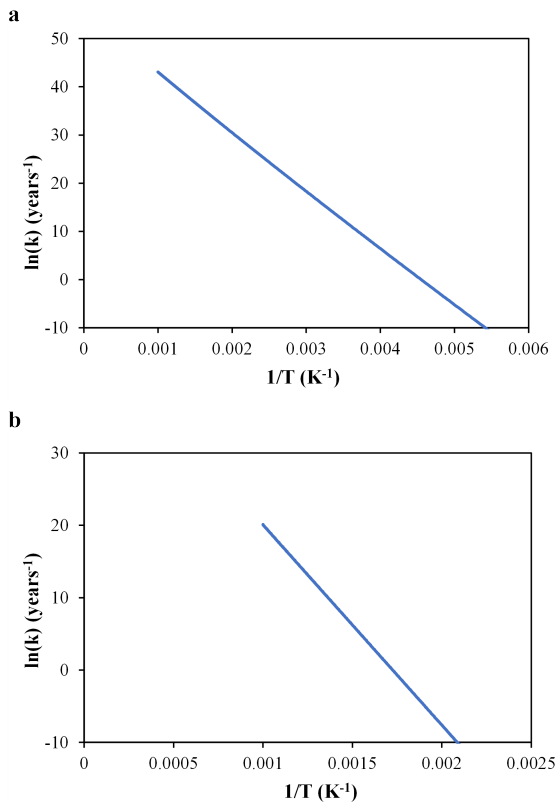

**Fig. 5.** Arrhenius plots for: (a) the free energy barrier associated with the $TS_{C-D}$ structure on the $Fe^0@SiO_2$ catalyst, and (b) the free energy barrier associated with the $TS_{C-D}$ structure on the $Fe^{2+}@SiO_2$ catalyst. Plots including the fits and correlation coefficients, and all values represented in the plots are collected in Figs. F.1 and F.2 and Tables F.1–F.4 and F.5–F.7.

agreement with the energy barriers obtained from the PES (55.07, 22.21, and 3.70 kcal mol$^{-1}$ for $TS_{C-D}$, $TS_{F-G}$, and $TS_{G-H}$, respectively). The kinetic data clearly indicate that the RDS of this process is the first step, in agreement with its highest energy barrier. Moreover, this step is endergonic, also ruling out tunnelling contributions. The process does not reach $k = 1$ yr$^{-1}$ until it reaches above 500 K. The third step (we note that the second step is considered barrierless) has a similar behaviour to the previous steps; it also presents a relatively high energy barrier and an endergonic character, although it is not as kinetically unfavourable (a rate constant around $k = 1$ yr$^{-1}$ is achieved at 250 K). In contrast, the last step presents a different kinetic behaviour due to its low energy barrier and exergonicity. Here tunnelling contributions are important, and the step is fast at 10–20 K.

Our kinetics results show that, for both FTT processes, the RDS is the first step, namely the first hydrogenation of CO from the first adsorbed $H_2$. This is consistent with the fact that this step is the one that presents the highest energy barrier in both reactions. However, on $Fe^0@SiO_2$, the RDS is significantly faster than on $Fe^{2+}@SiO_2$ since in the latter the energy barrier is more than twice that in the former. The reason is on the key cooperation between the silica surface and the Fe centre on $Fe^0@SiO_2$, which allows the spontaneous dissociation of $H_2$ when both reactants are adsorbed on the catalyst, a phenomenon that does not happen on $Fe^{2+}@SiO_2$.

According to this kinetic data, the "idoneous" catalytic systems to synthesise $CH_3OH$ through this FTT process in astrophysical environments should be those that are capable of easily dissociating $H_2$ and that keep the two H atoms close to CO, in which a small increase in temperature might trigger the reaction. Bare Fe metal grains (found in meteorites and in the resulting products of space weathering of Fe-containing silicates) would be suitable candidates.

Finally, it is worth mentioning that we propose here two different scenarios as regimes in which the FTT processes can operate (grains acting or not as third bodies), but these are two limiting situations, and accordingly intermediate situations can also take place. Determining whether the surface catalysts can dissipate the energies and in what degree can be theoretically elucidated by executing ab initio molecular dynamics (AIMD) simulations (see e.g. Pantaleone et al. 2020, 2021; Ferrero et al. 2023), which are computationally expensive and beyond the scope of this work.

### 4.2. Astrophysical implications: Fe catalysis in different environments

FTT reactions have been carried out in the laboratory under different astrophysical conditions. Hydrocarbons have been successfully synthesised using transition metal-containing dust analogues and under solar nebula conditions (Cabedo et al. 2021; Llorca & Casanova 1998; Ferrante et al. 2000; Sekine et al. 2006), but the formation of alcohols has been elusive. However, using meteoritic material and early Earth conditions, the formation of methanol and ethanol was achieved, although not until temperatures around 570 K and in lower percentages than hydrocarbons (Cabedo et al. 2021). Thus, it seems that methanol formation competes with the formation of hydrocarbons. However, formaldehyde was also detected as an FTT by-product, giving a consistency to the overall FTT picture since formaldehyde is a common FTT intermediate towards the subsequent formation of either hydrocarbons or alcohols. Accordingly, it seems that experimental formation of methanol could indeed occur through the FTT mechanism proposed here. Other indirect experimental evidence on the occurrence of FTT reactions forming hydrocarbons and alcohols is found by studying the composition and the petrology of carbonaceous chondrites (CCs) and interstellar dust particles (IDPs). Several studies found evidence of Fe carbides in IDPs (Mackinnon & Rietmeijer 1987; Bradley et al. 1984), which can be attributed to the carburisation of the Fe catalyst as a side reaction of FTT processes (Amelse et al. 1978; Unmuth et al. 1980). This is supported by the experimental production of Fe carbides during FTT under solar nebula conditions (Llorca & Casanova 1998). Brearley (2021) proposes that the presence of stable carbides only at high temperature is an evidence of catalytic carburisation in the solar nebula, or at least locally in the protoplanetary disks. FTT reactions in dust grains have fallen in disfavour through the years because they are unable to reproduce the right isotopic fractionation observed in chondritic meteorites (Johnson et al. 2012; Cronin & Pizzarello 1990), tending to the production of lighter isotopic ratios. However, since the amount of organic material formed by this mechanism is not dominant compared to other mechanisms, such as ice chemistry or parental body aqueous and/or thermal alteration, the distinct isotopic features of FTT might be hidden (Alexander et al. 2017).

Considering the plausible scenario where adsorption energies are dissipated by the grain surface, the energetics of the $Fe^0@SiO_2$ catalysis proposed here indicate that FTT reactions do not occur at temperatures lower than 150–200 K. This indicates that FTT-methanol formation is not expected to happen in very cold regions, such as dark clouds or pre-stellar cores where temperatures are around 10 K and dust grains are covered in volatile ices, in which potentially Fe catalytic inclusions are not exposed to the gas. However, as star and planet formation proceeds, different environments and energy sources are available, which can promote FTT reactions.

In protostellar cores, methanol is detected in the most extended envelopes (Kristensen et al. 2010) and in the most central regions, close to the protostars (Nazari et al. 2021): the hot cores or hot corinos (depending on the stellar mass). As mentioned in the Introduction, the large-scale presence of gas-phase methanol is explained by its formation from direct hydrogenation of CO ices and its subsequent desorption by non-thermal processes, such as grain-grain collisions or photon- and cosmic ray-induced desorption (Fuchs et al. 2009; Watanabe & Kouchi 2002). However, in hot corinos methanol can also undergo thermal desorption due to the rise in temperature (van Dishoeck et al. 1995; Van Der Tak 2000; Ceccarelli 2000; Schöier et al. 2002; Maret et al. 2005; Jorgensen et al. 2005). In these regions the bare dust grain surfaces are exposed to the gas-phase, which allows reactions; although the temperatures are high enough for FTT process to occur, the high abundance of atomic H makes the H-addition to CO the major channel leading to the formation of $CH_2O$ and $CH_3OH$, in detriment of FTT. However, FTT reactions can potentially also happen in other high-temperature environments, such as shocks or outflows, where methanol has also been detected (Bachiller et al. 1995; Bachiller & Pérez Gutiérrez 1997; Holdship et al. 2019). It is possible that in shock regions the methanol in the ice might be destroyed due to gas sputtering, and that the observed methanol is formed in post-shock reactions, perhaps via FTT (Suutarinen et al. 2014).

Protoplanetary disks are expected to inherit the methanol content from previous stellar formation stages. However, its detection has been elusive; it was only detected in one source (Walsh et al. 2016; Podio et al. 2019). Methanol appears confined to the disk midplane, predominantly at less than 100 AU from the central source, where it is expected to desorb from the icy grains. State-of-the-art astrochemical models help in reconciling the relative methanol high abundances, but the modelled radial distribution peaks are 2–3 orders higher than what is observed (Walsh et al. 2017). This suggests that even complex models miss a source of methanol in the innermost regions of the protoplanetary disks, which could well be FTT processes according to the inner disk conditions.

Methanol has been detected in comets (Hoban et al. 1991; Reuter 1992; Davies et al. 1993; Eberhardt' et al. 1994; Le Roy et al. 2015), which can arise from earlier stages of planetary evolution, including ices of primordial dust grains (Rubin et al. 2019). Thus, the methanol abundance is expected to vary according to the protoplanetary region in which it accreted (and therefore its origin), but also to the conditions under which it evolved, such as the protostellar gas and the protoplanetary disk temperatures, the CO depletion level in the primordial volatile gases (Faggi et al. 2023), or the irradiation conditions (Hudson & Moore 1999). While there is no direct evidence suggesting that FTT reactions take place during the cometary life, some of the organic components present in comets could have been inherited from FTT reactions occurring on grain surfaces in certain regions of the solar nebula (Llorca 2002), for example the presence of ethylene in cometary comas, whose origin is still elusive (Dartois 2021).

Finally, FTT reactions could be at work in the more advanced stages of planetary evolution, such as primitive atmospheres of terrestrial planets and/or planetary moons. Proto-Earth-sized protoplanets can efficiently accrete $H_2$ from the protoplanetary nebula (Stökl et al. 2016; Lammer et al. 2018), which is also rich in CO. In these environments, the presence of potentially catalytic metallic inclusions, for example by delivery of chondritic material, and the rising temperatures in planetary atmospheres constitute an ideal environment for FTT reactions, hence producing alcohols and hydrocarbons (Kress & McKay 2004).

## 5. Conclusions

Considering the cosmic high molecular $H_2$ and CO, and elemental Fe abundances, Fischer-Tropsch-type (FTT) reactions synthesising methanol using $Fe^0$ and $Fe^{2+}$ single-atom catalysis on silica surfaces ($Fe^0$@$SiO_2$ and $Fe^{2+}$@$SiO_2$) have been investigated here by means of quantum chemical simulations and kinetic calculations. The main conclusions of our study are the following:

1. A plausible reaction mechanism from which methanol is formed from the direct coupling of $H_2$ and CO through a heterogeneous catalytic process is described on both $Fe^0$@$SiO_2$ and $Fe^{2+}$@$SiO_2$ catalysts. However, the proposed mechanisms show different energetic barriers to overcome for the reaction to advance, and therefore the energetic requirements play a key role;

2. We demonstrate that in the scenario where the adsorption energies of $H_2$ and CO are not dissipated through the silica surface, the reaction catalysed by $Fe^0$@$SiO_2$ would advance in an apparent barrierless fashion as all the barriers are located below the asymptote. Moreover, calculated BEs demonstrate that the $CH_2O$ intermediate would remain adsorbed on the Fe centre, while the product $CH_3OH$ would desorb once formed. In contrast, the $Fe^{2+}$@$SiO_2$ process show a first barrier allocated above the asymptote meaning that an extra source of energy is needed in order for the reaction to evolve;

3. In the other scenario, where the adsorption energies are fully dissipated throughout the surface, our simulations indicate that FTT reactions on $Fe^0$@$SiO_2$ are, with energy requirements, affordable in different astrophysical environments. The key factor favouring the $Fe^0$@$SiO_2$ catalysis, in detriment of the $Fe^{2+}$@$SiO_2$ process, is the cooperation between the Fe centre and the silica surface, which helps a homolytic H-H cleavage in the pre-reactive adsorption complex (which does not happen on $Fe^{2+}$@$SiO_2$), in turn favouring the first CO hydrogenation step. Nevertheless, kinetic calculations indicate that this is the rate determining step, and accordingly hardly feasible in the deep cold regions of the ISM (i.e. dark molecular and pre-stellar cores). Moreover, the high abundance of atomic H in these regions and in protostellar hot cores and hot corinos suggest that, in these early astrophysical environments, $CH_3OH$ mainly forms via H addition to CO. Thus, the investigated FTT reactions require higher temperatures and higher $H_2$ abundances than those of the coldest interstellar regions and protostellar cores, meaning that its efficiency enhances in further stages of the planetary formation, with warmer temperatures, like protoplanetary disks, protoplanetary regions, and primitive planetary atmospheres. This is at variance with the primordial interstellar ice mantles, whose chemical activity is restricted to the first and cold stages of the planetary formation where temperatures keep them from sublimation;

4. In both scenarios FTT processes on $Fe^0$@$SiO_2$ are more affordable in terms of energy requirements, being those on

$Fe^{2+}$@$SiO_2$ almost impossible to be conceived in astrophysical conditions.

Our results thus demonstrate that true heterogeneous astrocatalysis is operative in diverse astrophysical environments, opening new and unexplored scenarios for the formation of complex C-containing molecules in space.

*Acknowledgements.* This project has received funding within the European Union's Horizon 2020 research and innovation program from the European Research Council (ERC) for the project "Quantum Chemistry on Interstellar Grains" (QUANTUMGRAIN), grant agreement no. 865657, and from the Marie Sklodowska-Curie for the project "Astro-Chemical Origins" (ACO), grant agreement no. 811312. G.P. thankfully acknowledges financial support by the Spanish Ministry of Universities and the European Union's Next Generation EU fund for a Margarita Salas contract. MICINN is also acknowledged for funding the project PID2021-126427NB-I00. The UK Engineering and Physical Science Research Council (EPSRC, EP/W023024/1) is also acknowledged. We also thankfully acknowledge the computer resources and assistance provided by the Barcelona Supercomputing Center (BSC) and CSUC.

## Appendix A: Benchmark study.

In order to choose the DFT functional that best represented the electronic behaviour of the system under study, we performed a benchmark study with a set of different DFT functionals. The reference values used here are the energies obtained at the CCSD(T) level (Raghavachari et al. 1989), and to perform these calculations at such a high level of theory we constructed a very small molecular cluster (Fig. A.1). The methodology followed for the benchmark study is the same as in the main work. The geometry of the molecular cluster was first optimised at the PBE-D3(BJ) level of theory; however, here this optimisation was performed both with the CP2K (Hutter et al. 2014) and Gaussian16 (Frisch et al. 2016) packages to prove that there were no significant differences between the two programs. The following single point calculations at the B3LYP-D3(BJ) (Becke 1993; Stephens et al. 1994), BHLYP-D3(BJ) (Becke 1993; Lee et al. 1988), M062X-D3 (Zhao & Truhlar 2004), MPWB1K-D3(BJ) (Zhao & Truhlar 2005), PW6B95-D3(BJ) (Zhao & Truhlar 2005), $\omega$B97X-D3 (Chai & Head-Gordon 2008), and CCSD(T) (Raghavachari et al. 1989) levels of theory were performed using the Gaussian16 package (Frisch et al. 2016). We note that in all cases the chosen basis set was a double-$\zeta$ basis set (DZVP), except for CCSD(T), which was the aug-cc-pVTZ basis set. Dispersion interactions were accounted for using Grimme's D3 correction or, if possible, the D3(BJ)version (Grimme et al. 2010). In Tables A and A.2 are gathered all the relative energies obtained from the different DFT functionals for the three spin states of the molecular cluster containing $Fe^{2+}$. We note that it was not possible to construct a similar molecular cluster for the $Fe^0$ system that was affordable in terms of time and computational costs at the CCSD(T) level. The results of the benchmark study pointed out that the B3LYP functional is the one that best describes the electronic character of the system under study as the relative energies for each state are those closest to the CCSD(T) values.

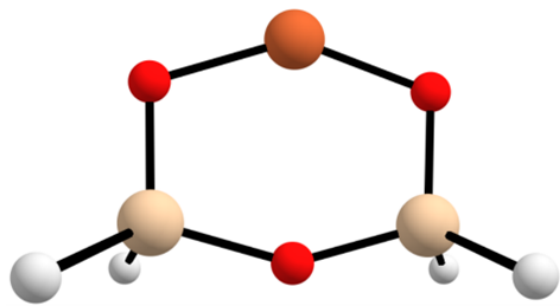

**Fig. A.1.** Optimised geometry for the molecular cluster used for the benchmark study. Colour-coding: white, H atoms; red, O atoms; beige, Si atoms; and orange, Fe atoms.

## Appendix B: Binding energies, comparison with experimental values.

For the sake of comparison, BEs between the reactants (adsorbates) CO and $H_2$ and the $SiO_2$ surface, where the CO and $H_2$ are either adsorbed at the Fe centre or on the bare surface, have been determined. These values have been compared with experimental data.

Binding energies with the metal centres are shown to be dramatically stronger than those with the bare silica surfaces. These values are expected as the type of interactions are different; the adsorbates are chemisorbed over the metal centre, while they are only physisorbed on the bare $SiO_2$ surfaces. Between the two different metal centres, $Fe^0$ demonstrates stronger interactions with the adsorbates than the $Fe^{2+}$. BEs for the bare silica surfaces are similar; it is worth noting that the surface with the lowest number of silanol groups show slightly higher binding energies. The interaction with the adsorbates in the first case happens directly on the surface, where there are more interacting points; however, in the later case those interactions happen directly on the OH groups of the silanol unities, thus reducing the number of interacting positions. Collings et al. (2015) performed a deep experimental study on the CO adsorption over $SiO_2$ surface models, where they concluded that binding energies are between 2.6 and 3 kcal mol$^{-1}$. Although there is a clear difference between the experimental results and computational results, both binding

**Table A.1.** Collected relative energies for the three spin states of the $Fe^{2+}$ atom in the $SiO_2$ molecular cluster.

| Structure | PBE-D3(BJ)[a] | PBE-D3(BJ)[b] | B3LYP-D3(BJ) | BHLYP-D3(BJ) | CCSD(T) |
|---|---|---|---|---|---|
| $Fe^{2+}@SiO_2$ (S) | 36.73 | 33.86 | 49.35 | 71.50 | 58.04 |
| $Fe^{2+}@SiO_2$ (T) | 23.51 | 21.58 | 38.29 | 55.81 | 43.68 |
| $Fe^{2+}@SiO_2$ (Q) | 0.00 | 0.00 | 0.00 | 0.00 | 0.00 |

**Notes.** [a]CP2K, [b]Gaussian16. Energies are in kcal mol$^{-1}$, spin states considered are singlet (S), triplet (T), and quintet (Q) multiplicities using as 0*th* reference energy the ground state (Q multiplicity). The energies are calculated at the PBE-D3(BJ), B3LYP-D3(BJ), BHLYP-D3(BJ), and CCSD(T) levels of theory.

**Table A.2.** Collected relative energies for the different spin states of the $Fe^{2+}$ atom in the $SiO_2$ molecular cluster.

| Structure | M062X-D3 | MPWB1K-D3(BJ) | PW6B95-D3(BJ) | $\omega$B97X-D3 | CCSD(T) |
|---|---|---|---|---|---|
| $Fe^{2+}@SiO_2$ (S) | 67.84 | 24.60 | 47.67 | 48.86 | 58.04 |
| $Fe^{2+}@SiO_2$ (T) | 49.98 | 17.59 | 38.44 | 56.29 | 43.68 |
| $Fe^{2+}@SiO_2$ (Q) | 0.00 | 0.00 | 0.00 | 0.00 | 0.00 |

**Notes.** Energies are in kcal mol$^{-1}$, spin states considered are singlet (S), triplet (T), and quintet (Q) multiplicities using as 0*th* reference energy the ground state (Q multiplicity). The energies are calculated at the M062X-D3, MPWB1K-D3(BJ), PW6B95-D3(BJ), $\omega$B97X-D3, and CCSD(T) levels of theory.

Table B.1. Collected binding energies (BEs).

| System | BE CO | BE $H_2$ |
|---|---|---|
| $Fe^0$@$SiO_2$ | 41.39 | 34.58 |
| $Fe^{2+}$@$SiO_2$ | 22.86 | 23.78 |
| $SiO_2$ (1.5 OH/$nm^2$) | 8.71 | 4.09 |
| $SiO_2$ (4.5 OH/$nm^2$) | 5.94 | 2.14 |

**Notes.** Energies are in kcal $mol^{-1}$; the BEs calculated here are between the adsorbates (CO and $H_2$) and each $SiO_2$ surface under study. Considering here both $Fe^0$@$SiO_2$ and $Fe^{2+}$@$SiO_2$ systems as well as the bare $SiO_2$ surfaces with each different silanol density.

energies affirms the physisorption of the CO molecule on the bare $SiO_2$ amorphous surface. What is clear from this data is the coordination between the CO and $H_2$ molecules and the Fe atoms.

## Appendix C: Relative adsorption energies

In the main text we consider only the adsorption of a $H_2$ molecule followed by the adsorption of a CO molecule. However, we also calculated the scenario where a CO molecule is first adsorbed and subsequently a $H_2$ molecule. In Table C.1 are collected all the relative energies for each of the possible adsorptions.

## Appendix D: Optimised geometries

In Figures D.1 and D.2 we show the optimised geometries for all the local minima and transition states studied in this work.

Table C.1. Collected relative internal energies ($\Delta U$).

| Surface | Reaction of adsorption | $\Delta U^0$ |
|---|---|---|
| $Fe^0$@$SiO_2$ | $H_2$ + $Fe^0$@$SiO_2$ → $H_2$/$Fe^0$@$SiO_2$ | -20.60 |
| | CO + $H_2$/$Fe^0$@$SiO_2$ → CO/$H_2$/$Fe^0$@$SiO_2$ | -60.41 |
| | CO + $Fe^0$@$SiO_2$ → CO/$Fe^0$@$SiO_2$ | -36.14 |
| | $H_2$ + CO/$Fe^0$@$SiO_2$ → $H_2$/CO/$Fe^0$@$SiO_2$ | -53.39 |
| $Fe^{2+}$@$SiO_2$ | H2 + $Fe^{2+}$@$SiO_2$ → H2/$Fe^{2+}$@$SiO_2$ | -17.69 |
| | CO + $H_2$/$Fe^{2+}$@$SiO_2$ → CO/$H_2$/$Fe^{2+}$@$SiO_2$ | -35.83 |
| | CO + $Fe^{2+}$@$SiO_2$ → CO/$Fe^{2+}$@$SiO_2$ | -21.70 |
| | $H_2$ + CO/$Fe^{2+}$@$SiO_2$ → H2/CO/$Fe^{2+}$@$SiO_2$ | -28.48 |

**Notes.** Values are in kcal $mol^{-1}$ for the adsorption of the reactants $H_2$ and CO for both systems under study, $Fe^0$@$SiO_2$ and $Fe^{2+}$@$SiO_2$, considering two scenarios, first the adsorption of the $H_2$ molecule followed by the adsorption of the CO molecule, and vice versa.

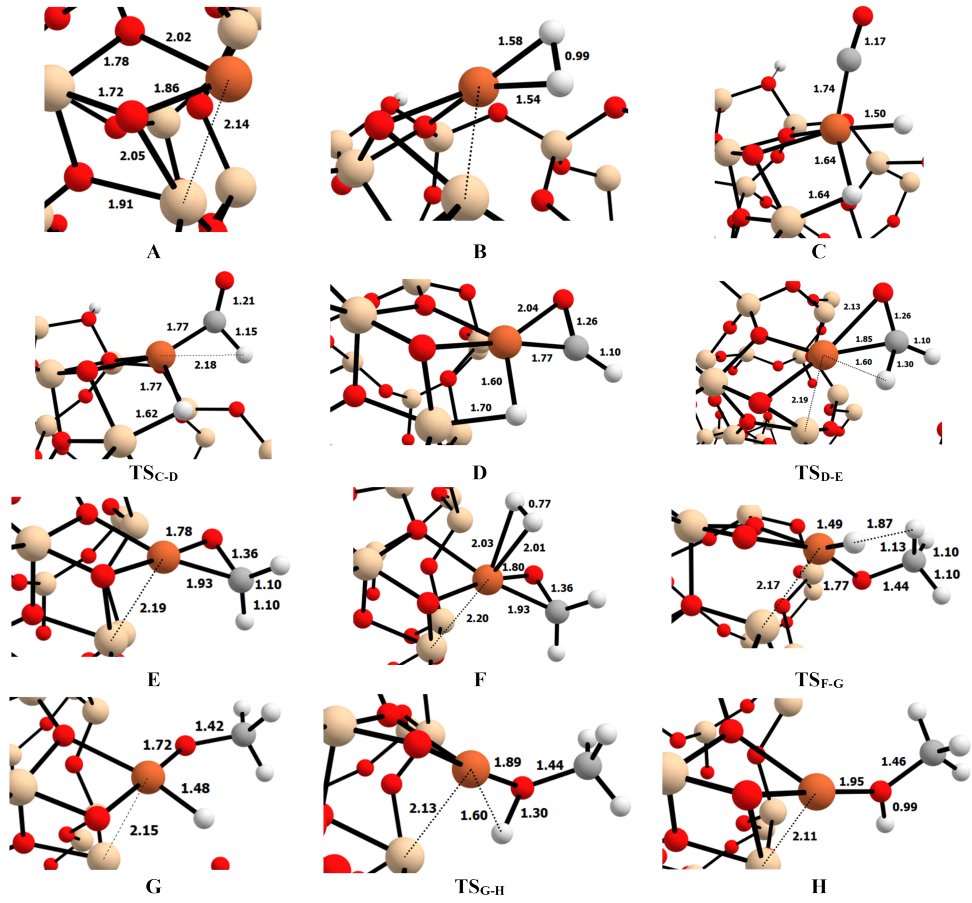

**Fig. D.1.** Optimised geometries for the all the local minima and transition states for the Fe$^0$@SiO$_2$ reaction process. Relevant distances depicted in angstroms Å. Colour-coding: white, H atoms; grey, C atoms; red O atoms; beige, Si atoms; and orange Fe atoms.

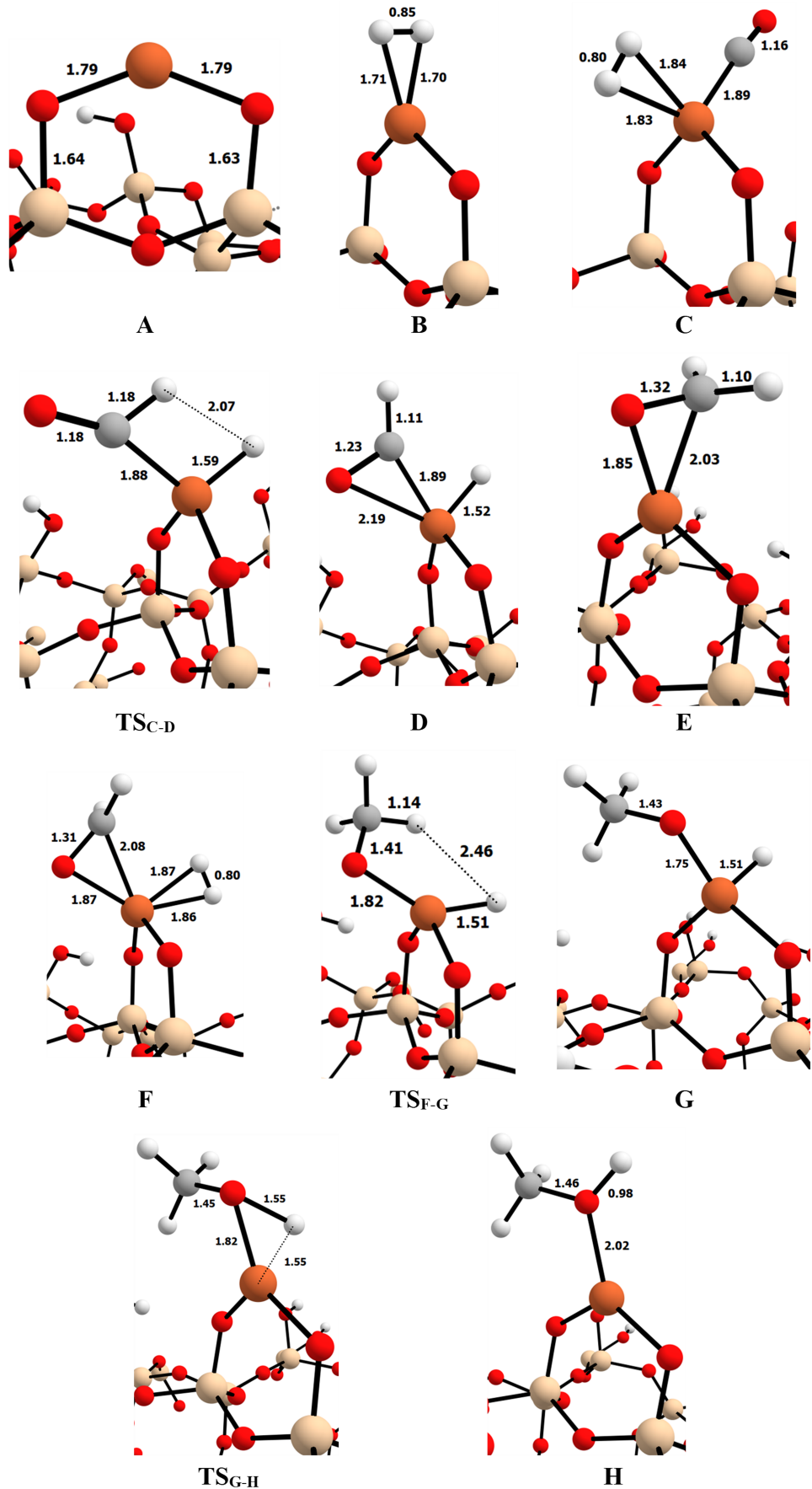

**Fig. D.2.** Optimised geometries for the local minima and transition states for the $Fe^{2+}@SiO_2$ reaction process. Relevant distances depicted in angstroms Å. Colour-coding: white, H atoms; grey, C atoms; red O atoms; beige, Si atoms; and orange Fe atoms.

## Appendix E: Relative reaction energies

From Table E.1 to Table E.4 are listed all the relative energies for both catalysts under study $Fe^0$@$SiO_2$ and $Fe^{2+}$@$SiO_2$, and for both levels of theory PBE-D3BJ/DZVP and B3LYP-D3BJ/TZVP.

**Table E.1.** Relative energies for the reaction mechanism $Fe^0$@$SiO_2$.

| System | ΔE | ΔZPE | ΔG (200 K) | ΔG (400 K) | ΔG (600 K) |
|---|---|---|---|---|---|
| A | 0.00 | 0.00 | 0.00 | 0.00 | 0.00 |
| B | -39.58 | -37.74 | -37.89 | -38.52 | -39.70 |
| C | -98.19 | -91.91 | -92.06 | -93.23 | -95.60 |
| $TS_{C-D}$ | -74.28 | -72.54 | -73.65 | -77.14 | -82.07 |
| D | -88.36 | -81.85 | -82.42 | -84.44 | -87.73 |
| $TS_{D-E}$ | -73.90 | -70.21 | -71.99 | -75.82 | -80.95 |
| E | -106.13 | -97.66 | -98.28 | -100.13 | -103.16 |
| F | -114.39 | -104.66 | -105.38 | -108.31 | -113.23 |
| $TS_{F-G}$ | -103.68 | -95.45 | -96.63 | -100.08 | -105.40 |
| G | -124.88 | -111.78 | -113.87 | -117.96 | -123.63 |
| $TS_{G-H}$ | -100.72 | -90.80 | -92.55 | -96.94 | -102.99 |
| H | -117.93 | -103.29 | -104.11 | -107.00 | -111.58 |

**Notes.** Energies are in kcal mol$^{-1}$; represented are the electronic energies (ΔE), zero-point corrected energies (ΔZPE), and Gibbs energies (ΔG) at 200, 300, and 600 K. The energies were obtained at the PBE-D3BJ/DZVP level of theory.

**Table E.2.** Relative energies for the reaction mechanism $Fe^0$@$SiO_2$.

| System | ΔE | ΔZPE | ΔG (200 K) | ΔG (400 K) | ΔG (600 K) |
|---|---|---|---|---|---|
| A | 0.00 | 0.00 | 0.00 | 0.00 | 0.00 |
| B | -22.44 | -20.60 | -20.76 | -21.38 | -22.56 |
| C | -66.69 | -60.41 | -60.56 | -61.72 | -64.10 |
| $TS_{C-D}$ | -40.52 | -38.78 | -39.89 | -43.38 | -48.31 |
| D | -52.16 | -45.65 | -46.22 | -48.24 | -51.53 |
| $TS_{D-E}$ | -34.18 | -30.48 | -32.27 | -36.10 | -41.23 |
| E | -69.99 | -61.52 | -62.15 | -64.00 | -67.02 |
| F | -90.02 | -80.29 | -81.01 | -83.94 | -88.85 |
| $TS_{F-G}$ | -75.75 | -67.51 | -68.69 | -72.15 | -77.46 |
| G | -100.02 | -86.92 | -89.02 | -93.10 | -98.78 |
| $TS_{G-H}$ | -79.84 | -69.92 | -71.66 | -76.06 | -82.10 |
| H | -107.04 | -92.39 | -93.21 | -96.11 | -100.68 |

**Notes.** Energies are in kcal mol$^{-1}$; represented are the electronic energies (ΔE), zero-point corrected energies (ΔZPE), and Gibbs energies (ΔG) at 200, 300, and 600 K. The energies were obtained at the B3LYP-D3BJ/TZVP level of theory.

**Table E.3.** Relative energies for the reaction mechanism $Fe^{2+}$@$SiO_2$.

| System | ΔE | ΔZPE | ΔG (200 K) | ΔG (400 K) | ΔG (600 K) |
|---|---|---|---|---|---|
| A | 0.00 | 0.00 | 0.00 | 0.00 | 0.00 |
| B | -20.20 | -18.11 | -18.51 | -19.66 | -21.39 |
| C | -61.43 | -55.98 | -56.34 | -58.17 | -61.45 |
| $TS_{C-D}$ | -18.15 | -14.27 | -14.76 | -16.59 | -19.64 |
| D | -40.53 | -33.98 | -34.93 | -37.21 | -40.73 |
| E | -72.90 | -64.45 | -65.01 | -66.88 | -69.95 |
| F | -83.51 | -73.22 | -73.99 | -76.81 | -81.52 |
| $TS_{F-G}$ | -69.60 | -58.60 | -59.53 | -62.33 | -66.64 |
| G | -86.39 | -73.93 | -74.92 | -77.95 | -82.63 |
| $TS_{G-H}$ | -76.05 | -65.69 | -66.94 | -70.56 | -75.82 |
| H | -101.75 | -86.81 | -88.00 | -91.44 | -96.54 |

**Notes.** Energies are in kcal mol$^{-1}$; represented are the electronic energies (ΔE), zero-point corrected energies (ΔZPE), and Gibbs energies (ΔG) at 200, 300, and 600 K. The energies were obtained at the PBE-D3BJ/DZVP level of theory.

**Table E.4.** Relative energies for the reaction mechanism $Fe^{2+}@SiO_2$.

| System | ΔE | ΔZPE | ΔG (200 K) | ΔG (400 K) | ΔG (600 K) |
|---|---|---|---|---|---|
| A | 0.00 | 0.00 | 0.00 | 0.00 | 0.00 |
| B | -19.78 | -17.69 | -18.09 | -19.24 | -20.98 |
| C | -41.28 | -35.83 | -36.19 | -38.02 | -41.30 |
| $TS_{C-D}$ | 15.16 | 19.04 | 18.54 | 16.72 | 13.67 |
| D | 2.85 | 9.40 | 8.45 | 6.17 | 2.65 |
| E | -49.93 | -41.48 | -42.04 | -43.92 | -46.98 |
| F | -52.27 | -41.98 | -42.75 | -45.57 | -50.28 |
| $TS_{F-G}$ | -30.85 | -19.85 | -20.78 | -23.58 | -27.89 |
| G | -50.39 | -37.93 | -38.92 | -41.95 | -46.63 |
| $TS_{G-H}$ | -44.64 | -34.28 | -35.52 | -39.15 | -44.41 |
| H | -87.59 | -72.65 | -73.84 | -77.28 | -82.38 |

**Notes.** Energies are in kcal mol$^{-1}$; represented are the electronic energies (ΔE), zero-point corrected energies (ΔZPE), and Gibbs energies (ΔG) at 200, 300, and 600 K. The energies were obtained at the B3LYP-D3BJ/TZVP level of theory.

## Appendix F: Kinetic data

In the main text we only depicted the Arrhenius plots for the rate determining steps of each reaction mechanism; however, the reaction constants for each energetic barrier were also calculated. In Figures F.1 and F.2 are listed the Arrhenius plots for all the considered barriers for the $Fe^0@SiO_2$ and $Fe^{2+}@SiO_2$ respectively. Moreover, from Table F.1 to Table F.7 are listed all the calculated values to construct the Arrhenius plots.

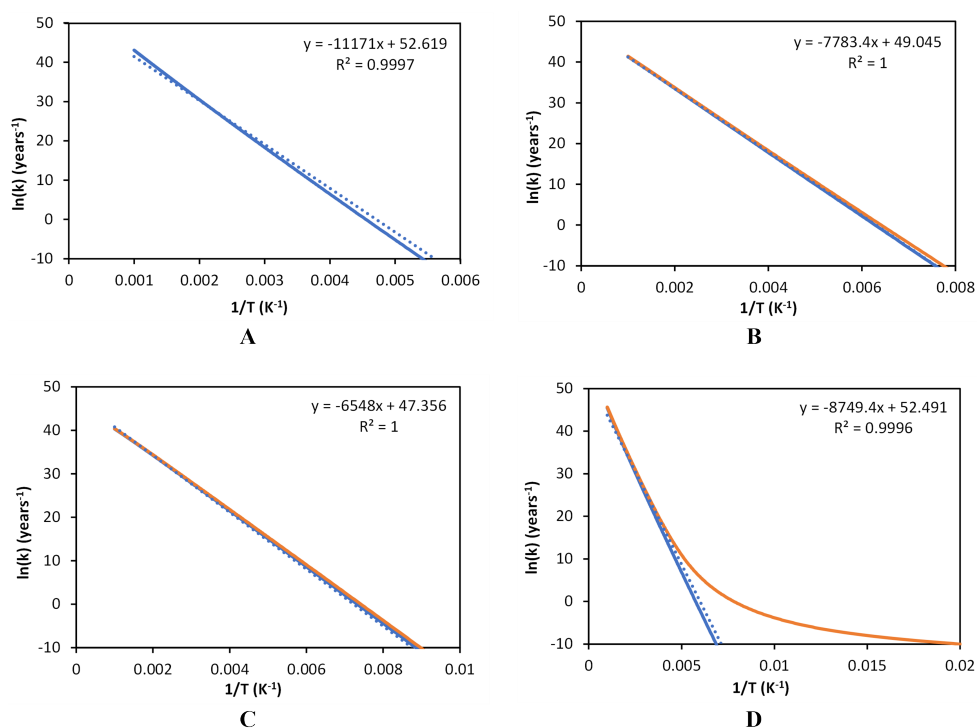

**Fig. F.1.** Arrhenius plots for $TS_{C-D}$, $TS_{D-E}$, $TS_{F-G}$, and $TS_{G-H}$ (A, B, C, and D, respectively) on the $Fe^0@SiO_2$ catalyst. Values of the natural logarithm of the rate constant (ln(k)) are represented in yr$^{-1}$ and the inverse of the temperatures (1/T) in K$^{-1}$. Represented in blue are the RRKM rate constants and in orange the rate constants considering tunnelling. Also included are the fits and correlation coefficients.

**Table F.1.** Kinetic data for the TS$_{C-D}$ on Fe$^0$@SiO$_2$.

| T (K) | k(T) (s$^{-1}$) | k(T) (y$^{-1}$) | 1/T (K$^{-1}$) | ln[k(T)] (s$^{-1}$) | ln[k(T)] (y$^{-1}$) |
|---|---|---|---|---|---|
| 50 | 4.98E-83 | 1.57E-75 | 2.00E-02 | -189.51 | -172.24 |
| 100 | 5.26E-35 | 1.66E-27 | 1.00E-02 | -78.93 | -61.66 |
| 150 | 9.31E-19 | 2.94E-11 | 6.67E-03 | -41.52 | -24.25 |
| 200 | 1.74E-10 | 5.50E-03 | 5.00E-03 | -22.47 | -5.20 |
| 250 | 1.99E-05 | 6.26E+02 | 4.00E-03 | -10.83 | 6.44 |
| 300 | 5.27E-02 | 1.66E+06 | 3.33E-03 | -2.94 | 14.32 |
| 350 | 1.58E+01 | 4.99E+08 | 2.86E-03 | 2.76 | 20.03 |
| 400 | 1.19E+03 | 3.76E+10 | 2.50E-03 | 7.08 | 24.35 |
| 450 | 3.55E+04 | 1.12E+12 | 2.22E-03 | 10.48 | 27.74 |
| 500 | 5.46E+05 | 1.72E+13 | 2.00E-03 | 13.21 | 30.48 |
| 550 | 5.19E+06 | 1.64E+14 | 1.82E-03 | 15.46 | 32.73 |
| 600 | 3.43E+07 | 1.08E+15 | 1.67E-03 | 17.35 | 34.62 |
| 650 | 1.71E+08 | 5.39E+15 | 1.54E-03 | 18.96 | 36.22 |
| 700 | 6.81E+08 | 2.15E+16 | 1.43E-03 | 20.34 | 37.60 |
| 750 | 2.27E+09 | 7.15E+16 | 1.33E-03 | 21.54 | 38.81 |
| 800 | 6.51E+09 | 2.05E+17 | 1.25E-03 | 22.60 | 39.86 |
| 850 | 1.66E+10 | 5.23E+17 | 1.18E-03 | 23.53 | 40.80 |
| 900 | 3.82E+10 | 1.20E+18 | 1.11E-03 | 24.37 | 41.63 |
| 950 | 8.06E+10 | 2.54E+18 | 1.05E-03 | 25.11 | 42.38 |
| 1000 | 1.58E+11 | 4.99E+18 | 1.00E-03 | 25.79 | 43.05 |

**Notes.** The columns give the rate constants (k(T)) in seconds (s$^{-1}$) and years (y$^{-1}$) for each 50 kelvin (K) and considering tunnelling (k(T)$_{Tun}$). Also shown is the inverse of the temperature (1/T) in K$^{-1}$ and the natural logarithm for each k(T) (ln[k(T)])).

**Table F.2.** Kinetic data for the TS$_{D-E}$ on Fe$^0$@SiO$_2$.

| T (K) | k(T) (s$^{-1}$) | k(T)$_{Tun}$ (s$^{-1}$) | k(T) (y$^{-1}$) | k(T)$_{Tun}$ (y$^{-1}$) | 1/T (K$^{-1}$) | ln[(k(T)] (s$^{-1}$) | ln[k(T)$_{Tun}$] (s$^{-1}$) | ln[(k(T)] (y$^{-1}$) | ln[k(T)$_{Tun}$] (y$^{-1}$) |
|---|---|---|---|---|---|---|---|---|---|
| 50 | 1.05E-54 | 1.80E-36 | 3.31E-47 | 5.69E-29 | 2.00E-02 | -124.29 | -82.30 | -107.03 | -65.04 |
| 100 | 7.81E-21 | 1.07E-19 | 2.46E-13 | 3.39E-12 | 1.00E-02 | -46.30 | -43.68 | -29.03 | -26.41 |
| 150 | 1.58E-09 | 4.60E-09 | 5.00E-02 | 1.45E-01 | 6.67E-03 | -20.26 | -19.20 | -3.00 | -1.93 |
| 200 | 6.89E-04 | 1.33E-03 | 2.17E+04 | 4.21E+04 | 5.00E-03 | -7.28 | -6.62 | 9.99 | 10.65 |
| 250 | 1.70E+00 | 2.73E+00 | 5.35E+07 | 8.60E+07 | 4.00E-03 | 0.53 | 1.00 | 17.80 | 18.27 |
| 300 | 3.17E+02 | 4.58E+02 | 9.98E+09 | 1.44E+10 | 3.33E-03 | 5.76 | 6.13 | 23.02 | 23.39 |
| 350 | 1.34E+04 | 1.82E+04 | 4.23E+11 | 5.73E+11 | 2.86E-03 | 9.50 | 9.81 | 26.77 | 27.07 |
| 400 | 2.24E+05 | 2.90E+05 | 7.08E+12 | 9.13E+12 | 2.50E-03 | 12.32 | 12.58 | 29.59 | 29.84 |
| 450 | 2.01E+06 | 2.51E+06 | 6.34E+13 | 7.90E+13 | 2.22E-03 | 14.51 | 14.73 | 31.78 | 32.00 |
| 500 | 1.16E+07 | 1.41E+07 | 3.66E+14 | 4.44E+14 | 2.00E-03 | 16.27 | 16.46 | 33.53 | 33.73 |
| 550 | 4.86E+07 | 5.78E+07 | 1.53E+15 | 1.82E+15 | 1.82E-03 | 17.70 | 17.87 | 34.97 | 35.14 |
| 600 | 1.60E+08 | 1.87E+08 | 5.05E+15 | 5.91E+15 | 1.67E-03 | 18.89 | 19.05 | 36.16 | 36.31 |
| 650 | 4.38E+08 | 5.05E+08 | 1.38E+16 | 1.59E+16 | 1.54E-03 | 19.90 | 20.04 | 37.16 | 37.31 |
| 700 | 1.04E+09 | 1.18E+09 | 3.27E+16 | 3.73E+16 | 1.43E-03 | 20.76 | 20.89 | 38.02 | 38.16 |
| 750 | 2.18E+09 | 2.46E+09 | 6.87E+16 | 7.76E+16 | 1.33E-03 | 21.50 | 21.62 | 38.77 | 38.89 |
| 800 | 4.17E+09 | 4.67E+09 | 1.31E+17 | 1.47E+17 | 1.25E-03 | 22.15 | 22.26 | 39.42 | 39.53 |
| 850 | 7.37E+09 | 8.19E+09 | 2.32E+17 | 2.58E+17 | 1.18E-03 | 22.72 | 22.83 | 39.99 | 40.09 |
| 900 | 1.22E+10 | 1.35E+10 | 3.85E+17 | 4.25E+17 | 1.11E-03 | 23.23 | 23.33 | 40.49 | 40.59 |
| 950 | 1.92E+10 | 2.10E+10 | 6.04E+17 | 6.63E+17 | 1.05E-03 | 23.68 | 23.77 | 40.94 | 41.04 |
| 1000 | 2.87E+10 | 3.13E+10 | 9.04E+17 | 9.88E+17 | 1.00E-03 | 24.08 | 24.17 | 41.35 | 41.43 |

**Notes.** The columns give the rate constants (k(T)) in seconds (s$^{-1}$) and years (y$^{-1}$) for each 50 kelvin (K) and considering tunnelling (k(T)$_{Tun}$). Also shown is the inverse of the temperature (1/T) in K$^{-1}$ and the natural logarithm for each k(T) (ln[k(T)])).

**Table F.3.** Kinetic data for the TS$_{F-G}$ on Fe$^0$@SiO$_2$.

| T (K) | k(T) (s$^{-1}$) | k(T)$_{Tun}$ (s$^{-1}$) | k(T) (y$^{-1}$) | k(T)$_{Tun}$ (y$^{-1}$) | 1/T (K$^{-1}$) | ln[k(T)] (s$^{-1}$) | ln[k(T)$_{Tun}$] (s$^{-1}$) | ln[k(T)] (y$^{-1}$) | ln[k(T)$_{Tun}$] (y$^{-1}$) |
|---|---|---|---|---|---|---|---|---|---|
| 50 | 1.94E-44 | 2.51E-43 | 6.11E-37 | 7.92E-36 | 2.00E-02 | -100.65 | -98.09 | -83.39 | -80.82 |
| 100 | 7.61E-16 | 1.95E-15 | 2.40E-08 | 6.16E-08 | 1.00E-02 | -34.81 | -33.87 | -17.55 | -16.60 |
| 150 | 2.27E-06 | 3.95E-06 | 7.15E+01 | 1.25E+02 | 6.67E-03 | -13.00 | -12.44 | 4.27 | 4.82 |
| 200 | 1.10E-01 | 1.63E-01 | 3.48E+06 | 5.13E+06 | 5.00E-03 | -2.20 | -1.82 | 15.06 | 15.45 |
| 250 | 6.92E+01 | 9.31E+01 | 2.18E+09 | 2.94E+09 | 4.00E-03 | 4.24 | 4.53 | 21.50 | 21.80 |
| 300 | 4.99E+03 | 6.35E+03 | 1.58E+11 | 2.00E+11 | 3.33E-03 | 8.52 | 8.76 | 25.78 | 26.02 |
| 350 | 1.05E+05 | 1.28E+05 | 3.31E+12 | 4.05E+12 | 2.86E-03 | 11.56 | 11.76 | 28.83 | 29.03 |
| 400 | 1.02E+06 | 1.21E+06 | 3.21E+13 | 3.82E+13 | 2.50E-03 | 13.83 | 14.01 | 31.10 | 31.27 |
| 450 | 5.90E+06 | 6.87E+06 | 1.86E+14 | 2.17E+14 | 2.22E-03 | 15.59 | 15.74 | 32.86 | 33.01 |
| 500 | 2.39E+07 | 2.73E+07 | 7.52E+14 | 8.62E+14 | 2.00E-03 | 16.99 | 17.12 | 34.25 | 34.39 |
| 550 | 7.43E+07 | 8.39E+07 | 2.34E+15 | 2.65E+15 | 1.82E-03 | 18.12 | 18.25 | 35.39 | 35.51 |
| 600 | 1.90E+08 | 2.12E+08 | 5.99E+15 | 6.70E+15 | 1.67E-03 | 19.06 | 19.17 | 36.33 | 36.44 |
| 650 | 4.18E+08 | 4.63E+08 | 1.32E+16 | 1.46E+16 | 1.54E-03 | 19.85 | 19.95 | 37.12 | 37.22 |
| 700 | 8.19E+08 | 8.99E+08 | 2.58E+16 | 2.84E+16 | 1.43E-03 | 20.52 | 20.62 | 37.79 | 37.88 |
| 750 | 1.46E+09 | 1.59E+09 | 4.60E+16 | 5.02E+16 | 1.33E-03 | 21.10 | 21.19 | 38.37 | 38.45 |
| 800 | 2.41E+09 | 2.61E+09 | 7.59E+16 | 8.24E+16 | 1.25E-03 | 21.60 | 21.68 | 38.87 | 38.95 |
| 850 | 3.73E+09 | 4.03E+09 | 1.18E+17 | 1.27E+17 | 1.18E-03 | 22.04 | 22.12 | 39.31 | 39.38 |
| 900 | 5.50E+09 | 5.91E+09 | 1.73E+17 | 1.86E+17 | 1.11E-03 | 22.43 | 22.50 | 39.69 | 39.77 |
| 950 | 7.76E+09 | 8.30E+09 | 2.45E+17 | 2.62E+17 | 1.05E-03 | 22.77 | 22.84 | 40.04 | 40.11 |
| 1000 | 1.05E+10 | 1.12E+10 | 3.32E+17 | 3.55E+17 | 1.00E-03 | 23.08 | 23.14 | 40.35 | 40.41 |

**Notes.** The columns give the rate constants (k(T)) in seconds (s$^{-1}$) and years (y$^{-1}$) for each 50 kelvin (K) and considering tunnelling (k(T)$_{Tun}$). Also shown is the inverse of the temperature (1/T) in K$^{-1}$ and the natural logarithm for each k(T) (ln[k(T)])).

**Table F.4.** Kinetic data for the TS$_{G-H}$ on Fe$^0$@SiO$_2$.

| T (K) | k(T) (s$^{-1}$) | k(T)$_{Tun}$ (s$^{-1}$) | k(T) (y$^{-1}$) | k(T)$_{Tun}$ (y$^{-1}$) | 1/T (K$^{-1}$) | ln[k(T)] (s$^{-1}$) | ln[k(T)$_{Tun}$] (s$^{-1}$) | ln[k(T)] (y$^{-1}$) | ln[k(T)$_{Tun}$] (y$^{-1}$) |
|---|---|---|---|---|---|---|---|---|---|
| 50 | 2.80E-62 | 1.40E-12 | 8.83E-55 | 4.43E-05 | 2.00E-02 | -141.73 | -27.29 | -124.46 | -10.02 |
| 100 | 1.31E-24 | 6.81E-10 | 4.13E-17 | 2.15E-02 | 1.00E-02 | -54.99 | -21.11 | -37.72 | -3.84 |
| 150 | 8.57E-12 | 7.89E-07 | 2.70E-04 | 2.49E+01 | 6.67E-03 | -25.48 | -14.05 | -8.22 | 3.21 |
| 200 | 3.02E-05 | 1.77E-03 | 9.51E+02 | 5.59E+04 | 5.00E-03 | -10.41 | -6.34 | 6.86 | 10.93 |
| 250 | 3.05E-01 | 2.18E+00 | 9.62E+06 | 6.87E+07 | 4.00E-03 | -1.19 | 0.78 | 16.08 | 18.05 |
| 300 | 1.58E+02 | 5.45E+02 | 4.99E+09 | 1.72E+10 | 3.33E-03 | 5.06 | 6.30 | 22.33 | 23.57 |
| 350 | 1.47E+04 | 3.59E+04 | 4.64E+11 | 1.13E+12 | 2.86E-03 | 9.60 | 10.49 | 26.86 | 27.75 |
| 400 | 4.59E+05 | 9.19E+05 | 1.45E+13 | 2.90E+13 | 2.50E-03 | 13.04 | 13.73 | 30.30 | 31.00 |
| 450 | 6.89E+06 | 1.21E+07 | 2.17E+14 | 3.83E+14 | 2.22E-03 | 15.75 | 16.31 | 33.01 | 33.58 |
| 500 | 6.16E+07 | 9.94E+07 | 1.94E+15 | 3.13E+15 | 2.00E-03 | 17.94 | 18.41 | 35.20 | 35.68 |
| 550 | 3.76E+08 | 5.69E+08 | 1.19E+16 | 1.79E+16 | 1.82E-03 | 19.75 | 20.16 | 37.01 | 37.43 |
| 600 | 1.72E+09 | 2.48E+09 | 5.43E+16 | 7.82E+16 | 1.67E-03 | 21.27 | 21.63 | 38.53 | 38.90 |
| 650 | 6.32E+09 | 8.74E+09 | 1.99E+17 | 2.76E+17 | 1.54E-03 | 22.57 | 22.89 | 39.83 | 40.16 |
| 700 | 1.94E+10 | 2.60E+10 | 6.12E+17 | 8.21E+17 | 1.43E-03 | 23.69 | 23.98 | 40.96 | 41.25 |
| 750 | 5.18E+10 | 6.77E+10 | 1.63E+18 | 2.13E+18 | 1.33E-03 | 24.67 | 24.94 | 41.94 | 42.20 |
| 800 | 1.23E+11 | 1.57E+11 | 3.88E+18 | 4.96E+18 | 1.25E-03 | 25.54 | 25.78 | 42.80 | 43.05 |
| 850 | 2.65E+11 | 3.33E+11 | 8.37E+18 | 1.05E+19 | 1.18E-03 | 26.30 | 26.53 | 43.57 | 43.80 |
| 900 | 5.28E+11 | 6.52E+11 | 1.66E+19 | 2.06E+19 | 1.11E-03 | 26.99 | 27.20 | 44.26 | 44.47 |
| 950 | 9.81E+11 | 1.20E+12 | 3.09E+19 | 3.77E+19 | 1.05E-03 | 27.61 | 27.81 | 44.88 | 45.08 |
| 1000 | 1.72E+12 | 2.07E+12 | 5.42E+19 | 6.53E+19 | 1.00E-03 | 28.17 | 28.36 | 45.44 | 45.63 |

**Notes.** The columns give the rate constants (k(T)) in seconds (s$^{-1}$) and years (y$^{-1}$) for each 50 kelvin (K) and considering tunnelling (k(T)$_{Tun}$). Also shown is the inverse of the temperature (1/T) in K$^{-1}$ and the natural logarithm for each k(T) (ln[k(T)])).

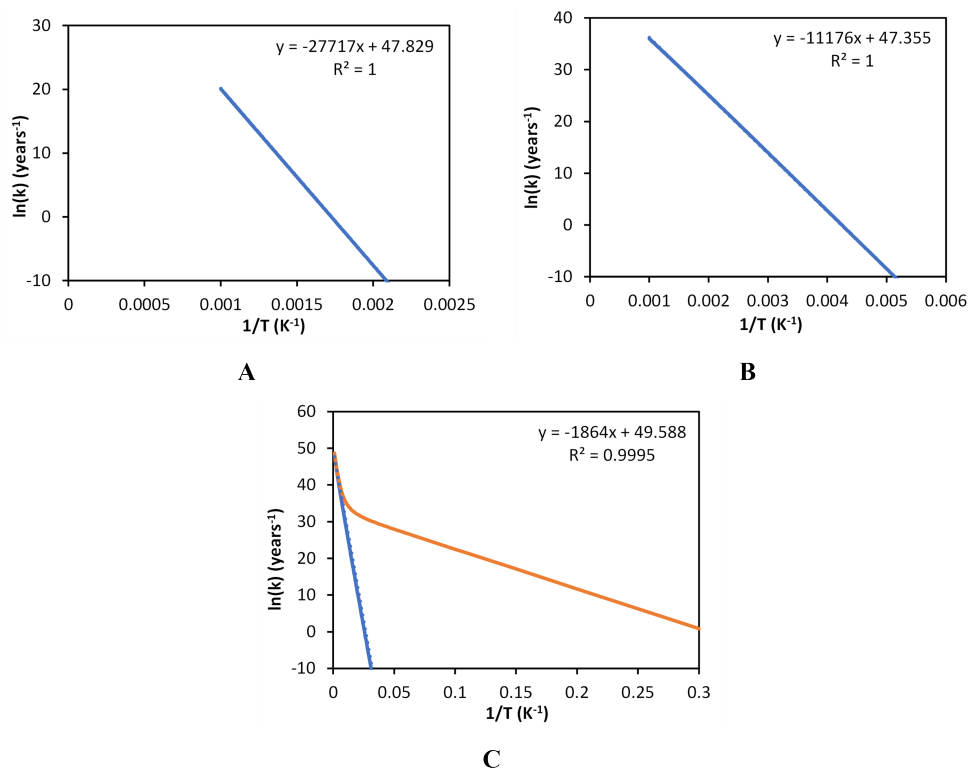

**Fig. F.2.** Arrhenius plots for TS$_{C-D}$, TS$_{F-G}$, and TS$_{G-H}$ (A, B, and C respectively) of the Fe$^{2+}$@SiO$_2$ catalyst. The values of the natural logarithm of the rate constant (ln(k)) are represented in years$^{-1}$ and the inverse of the temperatures (1/T) in kelvin (K). Represented in blue are the RRKM rate constants and in orange the rate constants considering tunnelling. Also included are the fits and correlation coefficients

**Table F.5.** Kinetic data for the TS$_{C-D}$ on Fe$^{2+}$@SiO$_2$

| T (K) | k(T) (s$^{-1}$) | k(T) (y$^{-1}$) | 1/T (K$^{-1}$) | ln[k(T)] (s$^{-1}$) | ln[k(T)] (y$^{-1}$) |
|---|---|---|---|---|---|
| 50   | 3.11E-228 | 9.80E-221 | 2.00E-02 | -523.86 | -506.59 |
| 100  | 7.82E-108 | 2.46E-100 | 1.00E-02 | -246.62 | -229.36 |
| 150  | 1.19E-67  | 3.74E-60  | 6.67E-03 | -154.10 | -136.84 |
| 200  | 1.35E-47  | 4.25E-40  | 5.00E-03 | -107.92 | -90.66  |
| 250  | 1.41E-35  | 4.45E-28  | 4.00E-03 | -80.25  | -62.98  |
| 300  | 1.46E-27  | 4.59E-20  | 3.33E-03 | -61.79  | -44.53  |
| 350  | 7.74E-22  | 2.44E-14  | 2.86E-03 | -48.61  | -31.34  |
| 400  | 1.53E-17  | 4.81E-10  | 2.50E-03 | -38.72  | -21.45  |
| 450  | 3.34E-14  | 1.05E-06  | 2.22E-03 | -31.03  | -13.76  |
| 500  | 1.57E-11  | 4.96E-04  | 2.00E-03 | -24.88  | -7.61   |
| 550  | 2.41E-09  | 7.61E-02  | 1.82E-03 | -19.84  | -2.58   |
| 600  | 1.60E-07  | 5.05E+00  | 1.67E-03 | -15.65  | 1.62    |
| 650  | 5.57E-06  | 1.76E+02  | 1.54E-03 | -12.10  | 5.17    |
| 700  | 1.17E-04  | 3.68E+03  | 1.43E-03 | -9.06   | 8.21    |
| 750  | 1.63E-03  | 5.14E+04  | 1.33E-03 | -6.42   | 10.85   |
| 800  | 1.64E-02  | 5.16E+05  | 1.25E-03 | -4.11   | 13.15   |
| 850  | 1.25E-01  | 3.96E+06  | 1.18E-03 | -2.08   | 15.19   |
| 900  | 7.67E-01  | 2.42E+07  | 1.11E-03 | -0.27   | 17.00   |
| 950  | 3.88E+00  | 1.22E+08  | 1.05E-03 | 1.35    | 18.62   |
| 1000 | 1.67E+01  | 5.25E+08  | 1.00E-03 | 2.81    | 20.08   |

**Notes.** The columns give the rate constants (k(T)) in seconds (s$^{-1}$) and years (y$^{-1}$) for each 50 kelvin (K) and considering tunnelling (k(T)$_{Tun}$). Also shown is the inverse of the temperature (1/T) in K$^{-1}$ and the natural logarithm for each k(T) (ln[k(T)]).

**Table F.6.** Kinetic data for the TS$_{F\text{-}G}$ on Fe$^{2+}$@SiO$_2$

| T (K) | k(T) (s$^{-1}$) | k(T) (y$^{-1}$) | 1/T (K$^{-1}$) | ln[k(T)] (s$^{-1}$) | ln[k(T)] (y$^{-1}$) |
|---|---|---|---|---|---|
| 50 | 7.49E-85 | 2.36E-77 | 2.00E-02 | -193.71 | -176.44 |
| 100 | 2.91E-36 | 9.19E-29 | 1.00E-02 | -81.82 | -64.56 |
| 150 | 5.13E-20 | 1.62E-12 | 6.67E-03 | -44.42 | -27.15 |
| 200 | 6.78E-12 | 2.14E-04 | 5.00E-03 | -25.72 | -8.45 |
| 250 | 5.07E-07 | 1.60E+01 | 4.00E-03 | -14.50 | 2.77 |
| 300 | 8.95E-04 | 2.82E+04 | 3.33E-03 | -7.02 | 10.25 |
| 350 | 1.85E-01 | 5.82E+06 | 2.86E-03 | -1.69 | 15.58 |
| 400 | 9.92E+00 | 3.13E+08 | 2.50E-03 | 2.29 | 19.56 |
| 450 | 2.17E+02 | 6.85E+09 | 2.22E-03 | 5.38 | 22.65 |
| 500 | 2.54E+03 | 8.01E+10 | 2.00E-03 | 7.84 | 25.11 |
| 550 | 1.88E+04 | 5.94E+11 | 1.82E-03 | 9.84 | 27.11 |
| 600 | 9.92E+04 | 3.13E+12 | 1.67E-03 | 11.51 | 28.77 |
| 650 | 4.03E+05 | 1.27E+13 | 1.54E-03 | 12.91 | 30.17 |
| 700 | 1.33E+06 | 4.20E+13 | 1.43E-03 | 14.10 | 31.37 |
| 750 | 3.74E+06 | 1.18E+14 | 1.33E-03 | 15.13 | 32.40 |
| 800 | 9.21E+06 | 2.90E+14 | 1.25E-03 | 16.04 | 33.30 |
| 850 | 2.03E+07 | 6.41E+14 | 1.18E-03 | 16.83 | 34.09 |
| 900 | 4.11E+07 | 1.30E+15 | 1.11E-03 | 17.53 | 34.80 |
| 950 | 7.69E+07 | 2.42E+15 | 1.05E-03 | 18.16 | 35.42 |
| 1000 | 1.35E+08 | 4.26E+15 | 1.00E-03 | 18.72 | 35.99 |

**Notes.** The columns give the rate constants (k(T)) in seconds (s$^{-1}$) and years (y$^{-1}$) for each 50 kelvin (K) and considering tunnelling (k(T)$_{Tun}$). Also shown is the inverse of the temperature (1/T) in K$^{-1}$ and the natural logarithm for each k(T) (ln[k(T)]).

**Table F.7.** Kinetic data for the TS$_{G\text{-}H}$ on Fe$^{2+}$@SiO$_2$.

| T (K) | k(T) (s$^{-1}$) | k(T)$_{Tun}$ (s$^{-1}$) | k(T) (y$^{-1}$) | k(T)$_{Tun}$ (y$^{-1}$) | 1/T (K$^{-1}$) | ln[k(T)] (s$^{-1}$) | ln[k(T)$_{Tun}$] (s$^{-1}$) | ln[k(T)] (y$^{-1}$) | ln[k(T)$_{Tun}$] (y$^{-1}$) |
|---|---|---|---|---|---|---|---|---|---|
| 50 | 8.99E-04 | 1.77E+06 | 4.01E+04 | 7.92E+13 | 2.00E-02 | -7.01 | 14.39 | 10.60 | 32.00 |
| 100 | 1.53E+05 | 5.95E+07 | 6.82E+12 | 2.66E+15 | 1.00E-02 | 11.94 | 17.90 | 29.55 | 35.52 |
| 150 | 1.05E+08 | 1.11E+09 | 4.69E+15 | 4.96E+16 | 6.67E-03 | 18.47 | 20.83 | 36.08 | 38.44 |
| 200 | 3.07E+09 | 1.08E+10 | 1.37E+17 | 4.85E+17 | 5.00E-03 | 21.85 | 23.11 | 39.46 | 40.72 |
| 250 | 2.52E+10 | 5.73E+10 | 1.13E+18 | 2.56E+18 | 4.00E-03 | 23.95 | 24.77 | 41.56 | 42.39 |
| 300 | 1.07E+11 | 1.96E+11 | 4.80E+18 | 8.75E+18 | 3.33E-03 | 25.40 | 26.00 | 43.01 | 43.62 |
| 350 | 3.12E+11 | 4.99E+11 | 1.39E+19 | 2.23E+19 | 2.86E-03 | 26.47 | 26.94 | 44.08 | 44.55 |
| 400 | 7.06E+11 | 1.04E+12 | 3.16E+19 | 4.64E+19 | 2.50E-03 | 27.28 | 27.67 | 44.90 | 45.28 |
| 450 | 1.35E+12 | 1.87E+12 | 6.04E+19 | 8.37E+19 | 2.22E-03 | 27.93 | 28.26 | 45.55 | 45.87 |
| 500 | 2.29E+12 | 3.04E+12 | 1.02E+20 | 1.36E+20 | 2.00E-03 | 28.46 | 28.74 | 46.08 | 46.36 |
| 550 | 3.55E+12 | 4.56E+12 | 1.59E+20 | 2.04E+20 | 1.82E-03 | 28.90 | 29.15 | 46.51 | 46.76 |
| 600 | 5.15E+12 | 6.43E+12 | 2.30E+20 | 2.87E+20 | 1.67E-03 | 29.27 | 29.49 | 46.88 | 47.11 |
| 650 | 7.07E+12 | 8.63E+12 | 3.16E+20 | 3.86E+20 | 1.54E-03 | 29.59 | 29.79 | 47.20 | 47.40 |
| 700 | 9.29E+12 | 1.12E+13 | 4.15E+20 | 4.99E+20 | 1.43E-03 | 29.86 | 30.04 | 47.48 | 47.66 |
| 750 | 1.18E+13 | 1.40E+13 | 5.28E+20 | 6.24E+20 | 1.33E-03 | 30.10 | 30.27 | 47.72 | 47.88 |
| 800 | 1.46E+13 | 1.71E+13 | 6.52E+20 | 7.62E+20 | 1.25E-03 | 30.31 | 30.47 | 47.93 | 48.08 |
| 850 | 1.76E+13 | 2.04E+13 | 7.87E+20 | 9.10E+20 | 1.18E-03 | 30.50 | 30.64 | 48.11 | 48.26 |
| 900 | 2.08E+13 | 2.39E+13 | 9.31E+20 | 1.07E+21 | 1.11E-03 | 30.67 | 30.80 | 48.28 | 48.42 |
| 950 | 2.43E+13 | 2.75E+13 | 1.08E+21 | 1.23E+21 | 1.05E-03 | 30.82 | 30.95 | 48.43 | 48.56 |
| 1000 | 2.78E+13 | 3.14E+13 | 1.24E+21 | 1.40E+21 | 1.00E-03 | 30.96 | 31.08 | 48.57 | 48.69 |

**Notes.** The columns give the rate constants (k(T)) in seconds (s$^{-1}$) and years (y$^{-1}$) for each 50 kelvin (K) and considering tunnelling (k(T)$_{Tun}$). Also shown is the inverse of the temperature (1/T) in K$^{-1}$ and the natural logarithm for each k(T) (ln[k(T)]).